\documentclass[twocolumn]{article}
\usepackage{comment}
\usepackage[english]{babel}
\usepackage{textgreek}
\usepackage[utf8]{inputenc}
\usepackage[colorinlistoftodos, color=green!40, prependcaption]{todonotes}
\usepackage{mathtools}
\usepackage{xcolor}
\usepackage{graphicx}
\usepackage[left=23mm,right=13mm,top=35mm,columnsep=15pt]{geometry} 
\usepackage{adjustbox}
\usepackage{placeins}
\usepackage[T1]{fontenc}
\usepackage{lipsum}
\usepackage{csquotes}
\usepackage{import}
\usepackage{xcolor}
\usepackage{hyperref}
\usepackage{pdflscape}
\usepackage{multicol}
\usepackage{caption}
\usepackage{appendix}

\title{Re-evaluating photoluminescent defects in Cu$_2$O}
\author{A. Brewin, M.P.A. Jones, S.J. Clark}
\date{\today}

\begin{document}

\maketitle

\begin{abstract}
    Defects in cuprous oxide (Cu$_2$O) strongly influence its performance in applications ranging from photovoltaics to emerging quantum technologies based on Rydberg excitons where microscopic crystal purity is essential. Photoluminescence (PL) spectroscopy is widely used as a diagnostic of material quality yet the origins of the sub-band-gap PL lines remain controversial despite decades of study. Using density functional theory we systematically evaluate native point defects in Cu$_2$O and identify which produce robust electronic states within the band gap. By combining supercell-size convergence criteria and cross-functional consistency checks we show that the widely accepted assignments of the 1.35\,eV, 1.5\,eV, and 1.7\,eV PL lines to copper and oxygen vacancies are unsupported. Instead we find that oxygen interstitials, copper interstitials and one form of split copper vacancy are the only native defects that consistently introduce true in-gap states. These results provide a revised framework for interpreting PL spectra and offer a more reliable basis for diagnosing crystal quality in Cu$_2$O-based quantum-device platforms.

\end{abstract}

\section{Introduction}


Cu$_2$O is one of the earliest semiconductors to be studied in detail, with a research history of over 70 years~\cite{bloem1958,apfel1960exciton}. In that time, it has been of interest for studying fundamental excitonic physics~\cite{apfel1960exciton}, solar cells~\cite{akimoto2006thin, chen2013review}, and most recently Rydberg excitons~\cite{Kazimierczuk2014}, where it is emerging as a platform for quantum sensing~\cite{heckotter2024rydberg}, polaritonic devices~\cite{orfanakis2022}, and microwave-optical conversion~\cite{pritchett2024,brewin2024}. For these quantum applications, microscopic control of the host crystal is crucial, since even small concentrations of defects can alter exciton properties such as lifetimes and coherence~\cite{Heckotter2020}. Indeed, sample variability is a central obstacle~\cite{Heckotter2020,Lynch2021}, where the exciton spectra from different samples of Cu$_2$O differ significantly in the heights of the absorption lines and the number of resolvable lines. This has been shown to be due to charged defects in the crystal~\cite{Kruger2020,bergen2023}, commonly taken to be copper and oxygen vacancies. The samples which show the highest principal quantum number exciton Rydberg states are all from natural crystals~\cite{Versteegh2021,bergen2023} and samples of such quality are rare. Work continues on the growth of synthetic cuprous oxide~\cite{steinhauer2020rydberg,mazanik2022strong,chang2013removal} to achieve reproducible material quality suitable for scalable quantum devices. The large sample‑to‑sample variability in Cu$_2$O spectroscopy suggests that native defects are neither rare nor simple, reinforcing the need for an assignment framework grounded in first‑principles calculations rather than historical conventions.

One method for assessing for sample quality is the photoluminescence (PL) spectrum~\cite{frazer2015,frazer2017,Lynch2021}. The PL spectrum in synthetic samples has been measured to contain significant emission of energy less than the band gap, $E_g$; emission that is not present in the spectra of high-quality natural samples~\cite{Lynch2021}. This is an indicator of defect states within the band gap due to defects in the crystal structure. For experiments using high‑n Rydberg excitons, even weak defect potentials lift degeneracies and broaden lines. Because PL lines probe the density and type of mid‑gap states correctly identifying their microscopic origin is essential for predicting and controlling the excitonic landscape in real samples.

\begin{figure}
    \centering
    \includegraphics[width=\linewidth]{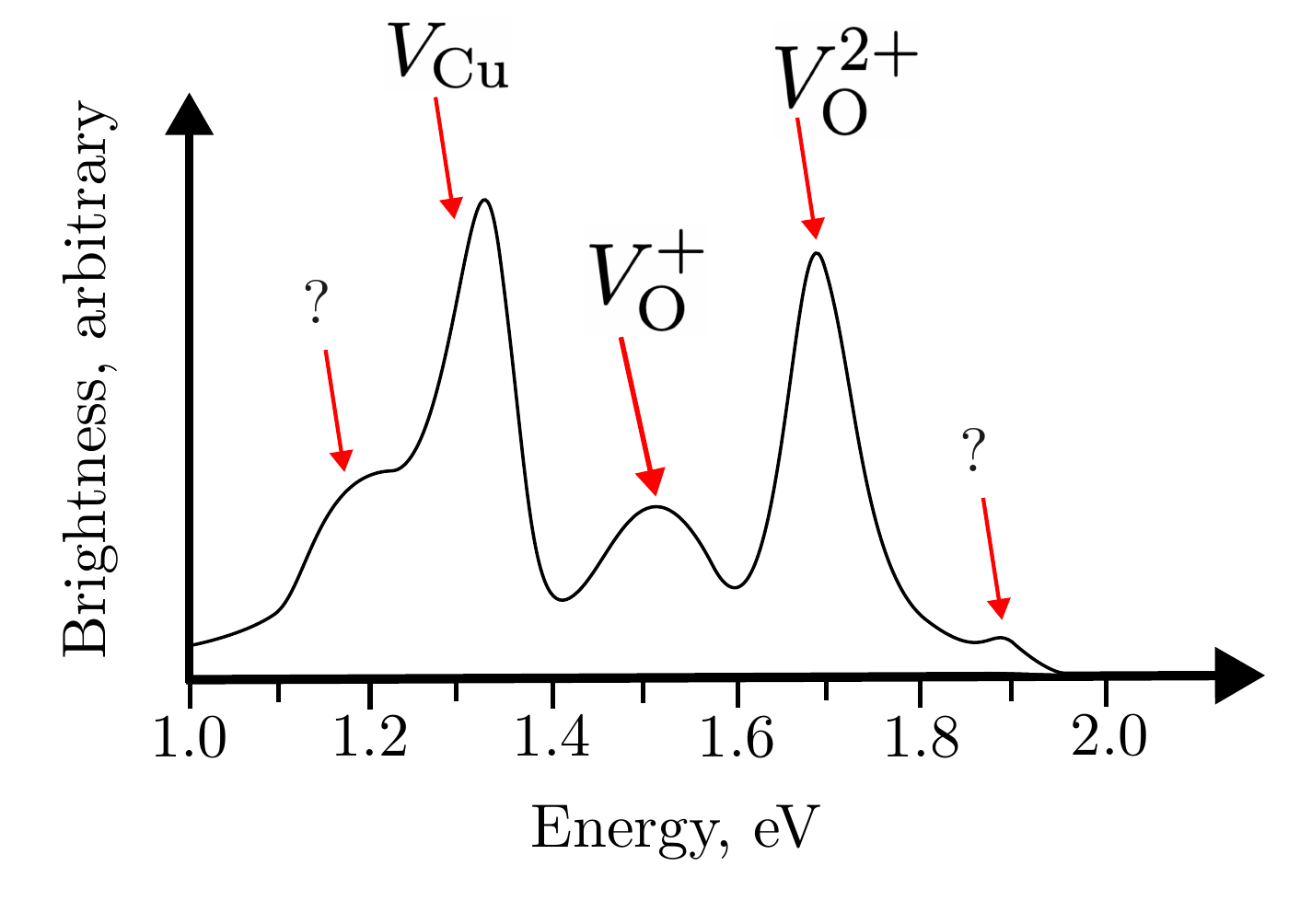}
    \caption{A sketch of an example PL spectrum showing the 4 main lines commonly observed, with the defect assignments made by Bloem~\cite{bloem1958}.}
    \label{fig:example_pl}
\end{figure}

The PL spectrum of Cu$_2$O has been studied many times over its history, and 5 main lines have been documented~\cite{bergen2023, Lynch2021, Kruger2020, frazer2017, frazer2015, koirala2013correlated, li2013engineering, ito1997,bloem1958}, at 1.2~eV, 1.35~eV, 1.5~eV, 1.7~eV, and 1.9~eV, as shown in figure \ref{fig:example_pl}. It is commonly stated in the field that the line at 1.35~eV is due to the copper vacancy, $V_\text{Cu}$, where a single copper atom is missing from the lattice. The lines at 1.5~eV and 1.7~eV are commonly attributed respectively to $V_\text{O}^+$ and $V_\text{O}^{2+}$, the different charge states of the oxygen vacancy. These assignments are due to a 1958 study by Bloem~\cite{bloem1958}, which linked changes in PL spectrum to the oxygen pressure during sample growth. In Bloem's samples the heights of the lines at 1.5~eV and 1.7~eV were always in the same ratio, and disappeared in the samples grown in high oxygen pressure. Under the assumption that the only significant native defects were $V_\text{Cu}$ and $V_\text{O}$, he therefore assigned the lines at 1.5~eV and 1.7~eV to $V_\text{O}^+$ and $V_\text{O}^{2+}$, and the large line at 1.35~eV to the remaining defect, $V_\text{Cu}$. However later studies showed that different samples displayed widely varying ratios or even absences of some lines, a discrepancy first noticed by Biccari~\cite{biccariThesis}. For example, the natural sample in Ito et al.~\cite{ito1997} showed a bright line for 1.7~eV and no line at all at 1.5~eV, suggesting the two lines were due to different defects. Furthermore Bloem did not speculate on the origin of the line at 1.2~eV, and the line at 1.9~eV has only been seen clearly in the recent results of Frazer et al.~\cite{frazer2017}. Finally, computational studies of Cu$_2$O~\cite{raebiger2007origins, soon2009, scanlon2009} using density functional theory (DFT) have shown that a host of native defects can form under various growth conditions, not just V$_\text{Cu}$ and V$_\text{O}$.

Despite the long history of PL measurements there remains little on a microscopically validated mapping between specific native defects and the observed PL lines. Earlier studies relied on indirect correlations or unconverged electronic‑structure calculations, leaving the field without a robust assignment scheme. In this work, we examine the native defects in Cu$_2$O from first principles using DFT. We identify which native point defects can, in principle, generate electronic states in the band gap, and therefore which are capable of producing the well known PL lines. Previous DFT studies of defects in Cu$_2$O identified up to 10 possible native defects~\cite{soon2009}, and found gap states associated with interstitials as well as vacancies~\cite{soon2009,scanlon2009}. However, more sophisticated calculations~\cite{isseroff2013} have shown these studies suffered from weak approximations and under-convergence with various DFT parameters. This leads to numerical artefacts - a problem that  Cu$_2$O is particularly sensitive to due to strong electron correlation effects~\cite{ruiz1997electronic,bruneval2006exchange}. Moreover, most previous calculations have been performed only in the minimal possible ($2\times2\times2$) supercell, which overestimates defect concentrations in the material and introduces finite-size effects, and only examine states at the gamma-point of the Brillouin zone, which can lead to mislabelling of gap states. These discrepancies highlight the need for a re-examination of the origin of these PL lines and motivates a systematic first‑principles survey designed specifically to eliminate supercell artefacts and misidentified folded bands.

Here we provide the converged, band‑structure‑level survey of all native point defects in Cu$_2$O, using both supercell scaling and hybrid‑functional cross‑validation to eliminate artefacts that have confounded earlier interpretations. And building on earlier studies, we examine all native point defects to investigate important finite-size effects; examine all defects with HSE-DFT, a more sophisticated approximation, to correct for band gap underestimation and explore exchange- and correlation-driven effects; analyse the full band structures of each defect and study the effect of charge on the band structures of the defect states.. Because defect‑induced levels can shift or appear artificially depending on supercell size, band‑folding effects and the choice of exchange–correlation functional, careful convergence testing is essential to avoid misassignment, allowing for a more nuanced categorisation of defect states.

\section{Methodology}
\subsection{Overview}
Our goal is to identify robust defect-induced in-gap states and so apply 
DFT, the leading theory for calculating the electronic structure of materials and molecules, to our defect systems. DFT is widely used to study the geometry, formation enthalpy, and electronic structure of both point defects and defect complexes in solids~\cite{samin2022analysis,li2023structural,smith2023structural,liu2021transition,murphy2020point,korkmaz2020defect}. While many experimental measurements are well-reproduced in DFT, such as atomic geometries, some are not usually quantitatively accurate, for example band gaps in semi-conductors. In fact, it is known that transition metal oxides in particular must be treated carefully with DFT~\cite{ostrom2022designing}. Nevertheless, choices of approximations can be made to mitigate these problems, and when used in a well understood way, can provide significant physical insight into defect-induced states. The novelty of our approach is to combine supercell-scaling, cross-functional verification (PBE vs HSE06) and a band-structure-level comparison against the pristine supercell. This aid the elimination of spurious in-gap features that can arise from band folding in small cells or from underconverged XC treatments.


When implementing DFT, the key choice to be made is of the treatment of exchange and correlation, via a term called the exchange-correlation (XC) functional. The choices vary in computational cost and accuracy for given observables. In this work, we use both the commonly used generalised gradient approximation of Perdew, Berke and Ernzerhof (PBE)~\cite{perdew1996}, as well as the hybrid functional of Heyd, Scuseria and Ernzerhof (HSE06)~\cite{heyd2003,heyd2004efficient}. The former is computationally much more efficient, and reproduces well geometries and formation energies. However, PBE significantly underestimates band gaps, which could present an issue in identifying mid-gap defect states. The latter  (HSE06) is an XC-functional designed to reproduce the band gaps of semiconductors much more accurately. It entails significantly higher computational cost, so we were only able to employ it for $2\times2\times2$ supercells. Our approach is to search for defect states that are robust to increases in supercell size, for which we use PBE, and robust to a more complex treatment of exchange and correlation, for which we use HSE.

The plane wave DFT code CASTEP~\cite{clark2005first} was used to run all the calculations in this work. We used pseudopotentials for the copper and oxygen ions generated using CASTEP's on-the-fly pseudopotentials. Oyxgen atoms had a core of helium and copper atoms had a core of argon, and in both cases PBE was used to compute the atomic core. 

Calculations with the Cu$_2$O unit cell under PBE and ultrasoft pseudopotentials found a relaxed cell size of 4.29~\AA, which was the lattice parameter used for all calculations, including those with HSE06. The optimisation of the cell size was considered converged when the change in energy per atom, $dE/d(\text{atom}) < 2\times10^{-5}$~eV, and the stress was below 0.1~GPa. For the same calculation, the density of the Monkhorst-Pack (MP) $k$-point grid and the cutoff energy for the plane wave basis were varied. The total energy was found to be converged to within 1~meV with an MP grid of $7\times7\times7$ and a plane wave cutoff energy of 1100~eV. Due to the inverse relation between real-space and k-space, the $2 \times 2 \times 2$ and $3\times3\times3$ supercell calculations needed only MP k-point grids of $4\times4\times4$ and $2\times2\times2$ to be similarly converged. The atomic positions for the various defect supercells were also found via geometry optimisation. These optimisations were considered converged when $dE/d(\text{atom}) < 2\times10^{-5}$~eV and all forces on atoms were less than $5\times10^{-2}$~eV/\AA.

For the HSE06 calculations, due the available computational resources, we used a cutoff of $E_\text{cut}=1000$~eV, as well as a reduced k-point grid sampling of $4\times 4\times 4$. This converged the energy difference between the valence band maximum (VBM) and the conduction band minimum (CBM) to within 1~meV, indicating (but unfortunately not guaranteeing) that the band structure has stopped changing in a material way.

The HSE functional contains two parameters that can be adjusted - the Hartree-Fock short-range exchange fraction, $\alpha$, and screening length $\omega $. We choose to set these at the default values of $\alpha = 0.25$, and $\omega = 0.2$~\AA$^{-1}$ determined from perturbation theory and fits to many materials~\cite{heyd2004efficient,peralta2006}. Scanlon et al.~\cite{scanlon2009} obtained a more accurate HSE-DFT band gap of $E_g^\text{HSE} = 2.12$~eV by changing the fraction of HF exchange from $\alpha = 0.25$ to $\alpha = 0.275$. However, given the systemic problems DFT often has with band gaps, there is no guarantee that a more accurate band gap will lead to better accuracy in any other property of the material. For this reason, we choose to keep the mixing fraction and screening length at their default values.


\subsection{Finite size effects}
Experimentally, the fractional deviation from stoichiometry in a sample is of the order $10^{-5}$~\cite{biccariThesis}, which is equivalent to 1 point defect per $\sim 10^4$ unit cells. This level of dilution is not achievable with the resources available for this project. However, for the correct physics, the defects need only be dilute enough to not `feel' the effects of neighbouring defects. To investigate where this dilute limit lies, we model each type of defect under PBE in a $2\times 2\times 2$ and $3\times 3\times 3$ supercell. As we will show, the decay of finite-size effects can be inferred from the difference between the PBE results for the two sizes of supercell with and without the defect, and varies significantly between defects.  Since HSE06 is a hybrid functional, it is much more expensive to compute, and so we did not run calculations under HSE06 for $3\times 3\times 3$ supercells. However the convergence of the PBE band structure with supercell size can often be used to infer likely impact of finite-size effects on the HSE calculations. We deem a defect level “finite-size-safe’’ if its energy relative to the pristine VBM shifts insignificantly upon going from $2\times2\times2$ to $3\times3\times3$ and its dispersion across the sampled Brillouin zone also remains negligible. Levels that fail either criterion are treated as finite-size artefacts or folded bulk bands and are discarded.

\subsection{Spin-orbit coupling}
Spin-orbit coupling (SOC) is an important effect in pure Cu$_2$O, especially for excitons, and is responsible for the various higher energy exciton series above the lowest-lying ``yellow'' series~\cite{schweiner2017even} through the splitting of the VBM. However, the numerical difference it makes to band energies is often small, of the order $0.1$~eV, and its inclusion introduces significant computational costs, as well as convergence issues associated with a non-collinear spin treatment. Given its small expected impact on defect state energies, it was not included in the results in this work. In addition, the experimentally relevant shifts for the PL-assigned levels exceed this scale, so omitting SOC in the largest cells does not affect our qualitative conclusions.

To understand the effect of this choice, SOC calculations were performed for the $2\times2\times2$ defect supercells under PBE, where it was introduced through CASTEP's spin-orbit coupling pseudopotentials, with the quantisation axis $(0,0,1)$. Examples of the results are compared to those calculations without SOC in figure \ref{fig:soc} in the appendix. Across the suite of defects, the inclusion of SOC did not affect the existence or non-existence of a defect state in the band gap; at most, it changed the energy gap of the defect state above the VBM by 0.2~eV. Therefore, we consider there to be a $\pm0.2$~eV error on the position of the defect states within the band gap due to spin-orbit coupling.

\subsection{Charge}
Charged point defects are believed to play an important role in modifying the excitonic properties of Cu$_2$O particularly in the context of recent studies on Rydberg excitons interacting with impurity potentials~\cite{Kruger2020,bergen2023}. We have investigated how changing the charge state of a defect affects the presence and energy of any in‑gap defect levels. Our primary goal here is to determine whether charging qualitatively alters the band structures and the positions of defect levels. We therefore focus on qualitative trends (appearance/disappearance and monotonic shifts of levels) rather than precise charge-transition levels, which would require larger cells and dielectric/finite-size corrections beyond the present scope. We restrict charged‑defect calculations to the PBE functional in the $3\times3\times3$ supercell, where finite‑size artefacts are reasonably mitigated and the PBE functional will give us the required qualitative results with the aim to assess trends in the appearance of in‑gap states rather than the absolute charge‑transition levels.

For each defect studied in the $3\times3\times3$ PBE supercell, we examined charge states in the range $q=-2,...,+2$. Structural relaxations were performed for each charged configuration to ensure that any charge‑induced local distortions were properly captured. In most cases, particularly for the simple copper and oxygen vacancies, the oxygen anti-site and the second split vacancy, changing the charge state did not introduce new defect levels in the band gap. The band structures differed from their neutral counterparts only by the expected occupation or depletion of already‑present states, indicating that none of these defects acquire an in‑gap state upon charging. In contrast, some defects that already show in‑gap states in the neutral configuration, such as the oxygen interstitials, exhibit systematic shifts in the energies of these states upon charging. For example, the defect levels associated with O interstitials shift rigidly upward by approximately 0.1–0.2 eV per added electron, consistent with the localised nature of the defect orbitals. This behaviour suggests that different charge states of the same defect could give rise to multiple nearby photoluminescence lines in experiment. By comparison, defects whose electronic states are strongly delocalised, such as the copper interstitials, display only small (~0.05 eV per electron) charge‑induced shifts reflecting their extended real-space character. Overall, the PBE calculations indicate that charging does not generate mid-gap states for defects that lack them in the neutral configuration, but can modify the relative energies of pre‑existing localised states. Although the quantitative values of the charged-defect energies are approximate, the qualitative trends identified here form the basis for the charge‑dependent defect-level shifts discussed in Sec. 6.

\subsection{Formation Enthalpies}

To understand which defects are likely to be present in real crystals (and therefore visible in PL) we compute the formation enthalpy, $\Delta H_\text{F}(D,q)$, of a defect $D$ with charge $q$. A low formation enthalpy means the defect can appear in meaningful concentrations under typical growth conditions, whereas a high formation enthalpy means the defect will be rare, and therefore is unlikely to produce a PL signal even if it produces a defect state in the band gap. In general, it is given by
\begin{align}
    \Delta H_\text{F}(D,q) = E(D,q) &- E_\text{P} + \sum_i n_i(E^\text{elem}_i + \Delta\mu_i) \nonumber\\
    &+ q[\epsilon_\text{VBM}^\text{P} + \Delta E_\text{F}]  + \Delta v(D,q),
\end{align} where $E(D,q)$ is the energy of the supercell with defect $D$ and charge $q$, $E_P$ is the energy of the neutral perfect supercell of the same size, and $n_i$ is the number of added (negative $n_i$) or subtracted (positive $n_i$) atoms of species $i$. $E^\text{elem}_i$ is the total energy per atom of each species in its reference state, in this case O$_2$(g) and Cu(s). The reasoning behind the first 3 terms is simply that the formation enthalpy is the energy of the final state minus the energy of the initial state, where the Cu and O atoms are assumed to come from or return to their reference states. The additional term $\Delta \mu_i$ changes the chemical potentials of each species depending on the particular chemical conditions in the furnace, whose allowed values we discuss below.

To calculate the cost of charging the defect, we not only need the energy of the charged supercell, but also the chemical potential of the electron(s) added or subtracted, whose value we take to be $\epsilon_\text{VBM}^\text{P}$, the eigenenergy of the VBM in the pure supercell at $\Gamma$. $\Delta E_\text{F}$ represents a hypothetical change in the Fermi level, for example from an applied voltage, which is set to zero at the VBM. 

When performing calculations on charged cells, due to periodic boundary conditions, we must introduce a uniform background potential of the opposite charge in order to keep the energies finite. Therefore, to make sure this does not affect the formation enthalpy calculations, we realign the energy of the charged cell with that of the neutral bulk by aligning the energy of an oxygen 2s orbital far away from the defect, via the term $\Delta v(D,q)$. It is defined here as the difference between two low-lying oxygen 2s energy levels in the neutral perfect and charged defect supercells, $\Delta v(D,q) = \Tilde{\epsilon}_\text{2s}^\text{P} - \Tilde{\epsilon}_\text{2s}^\text{far}(D,q)$, where $\Tilde{\epsilon}$ represents averaging the energy over the Brillouin zone. The 2s level used for the defect supercell is on an oxygen atom far from the defect, to ensure it is most comparable to the energy level in the bulk.



The chemical potentials, $\Delta \mu_i$, which represent the growth environment, are subject to several inequalities to prevent precipitates forming in the furnace. To avoid precipitation of O$_2$ and Cu metal,
\begin{align}
    \Delta \mu_\text{Cu} &\leq 0,\text{ and}\\
    \Delta \mu_\text{O} &\leq 0.
\end{align} Likewise, to avoid precipitation of CuO, it must be kept such that
\begin{align}
    \Delta \mu_\text{Cu} + \Delta \mu_\text{O} \leq \Delta H_\text{F}(\text{CuO}).
\end{align} Finally, in equilibrium when forming Cu$_2$O it must be true that
\begin{align}
\label{eq:hfcu2o}
    2\Delta \mu_\text{Cu} + \Delta \mu_\text{O} = \Delta H_\text{F}(\text{Cu$_2$O}).
\end{align} Along the line defined by equation \ref{eq:hfcu2o}, the endpoints correspond to the physically important extremes (Cu rich/O poor and Cu poor/O rich) which bracket realistic growth conditions.
\begin{center}
    \centering
    \begin{tabular}{ c c c }
        & Cu rich/O poor & Cu poor/O rich\\
        \hline
        $\Delta\mu_\text{Cu}$, eV & 0 & -0.33 \\
        $\Delta\mu_\text{O}$, eV & -1.80 & -1.30 \\ 
    \end{tabular}
    \label{table:dmu}
\end{center}
The elemental chemical potentials $\mu^\text{elem}_\text{Cu}$ and $\mu^\text{elem}_\text{O}$ are obtained from the energy per atom of the `natural' state of the elements, in this case O$_2$ and metallic Cu. For consistency, their geometries were first optimised under PBE, giving a lattice constant of $3.63$~\AA~for Cu and a bond length of $1.21$~nm for O$_2$, resulting in $E^\text{elem}_\text{Cu}=1680.93$~eV/atom and $E^\text{elem}_\text{O}=436.80$~eV/atom.

\subsection{Criteria for a defect state}
\label{sec:criteria}
Identifying a genuine defect‑induced electronic state within the Cu$_2$O band gap requires care, given the well‑known artefacts that arise in supercell calculations. Supercells inevitably introduce folded host bands and additional states unrelated to the defect, while finite-size interactions can shift or distort the energies of the true defect levels. To avoid misinterpretation, we adopt a criterion: a state is classified as a genuine defect level only if it consistently appears between the bulk valence‑band maximum (VBM) and the bulk conduction‑band minimum (CBM) across both combinations of exchange–correlation functional (PBE and HSE06) and supercell size, and if its presence cannot be attributed to band folding or to perturbations arising from artificially high defect concentrations. This approach ensures that only those states that persist under both changes in the computational method and dilution of the defect are retained as physically meaningful. 


To apply this criterion reliably, each defect supercell band structure is compared directly with the corresponding pure supercell band structure. This comparison allows us to distinguish genuinely new states induced by the defect from bands that arise simply due to the modified Brillouin zone or the additional electrons in the supercell. States that remain nearly dispersion-less across the Brillouin zone, shift little with increasing supercell size, and remain separated from folded bulk states provide strong evidence of real-space localisation and are therefore likely to survive in the dilute‑defect limit. Conversely, states that are strongly dispersive, appear only in smaller supercells, or shift appreciably when changing the XC functional or supercell size are treated as artefacts of finite-size effects or band folding and are not classified as true defect states. Applying this methodology, only the oxygen interstitials satisfy the full set of criteria, producing well‑defined in‑gap states for all tested functionals and supercell sizes. Other defects occasionally produce states that appear promising at a single level of theory but fail the consistency test; these are designated as ambiguous cases and are discussed separately.


\section{DFT on pure Cu$_2$O}

We first establish the baseline electronic structure of pristine Cu$_2$O, which serves as the reference against which all defect-induced changes are identified. We then present a defect-by-defect analysis, applying the robustness criteria to determine which native defects generate genuine in-gap states and which do not. This structure allows direct comparison between pure and defective supercells and ensures that artefacts arising from band folding or unconverged supercell sizes are excluded.
This serves two purposes: first, it provides reference band structures to compare with defect supercells; and second, it allows us to determine which of the features of the pure band structure are robust with respect to supercell enlargement. This will become crucial when identifying spurious states.  

\begin{figure}
    \centering
    \includegraphics[width=0.9\linewidth]{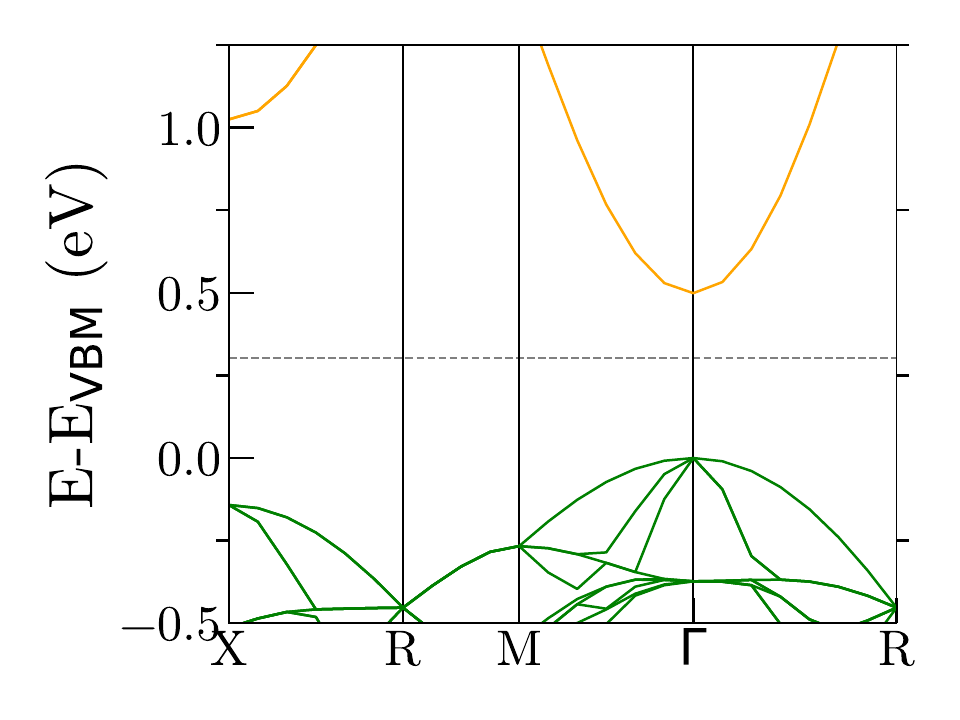}
    \includegraphics[width=0.9\linewidth]{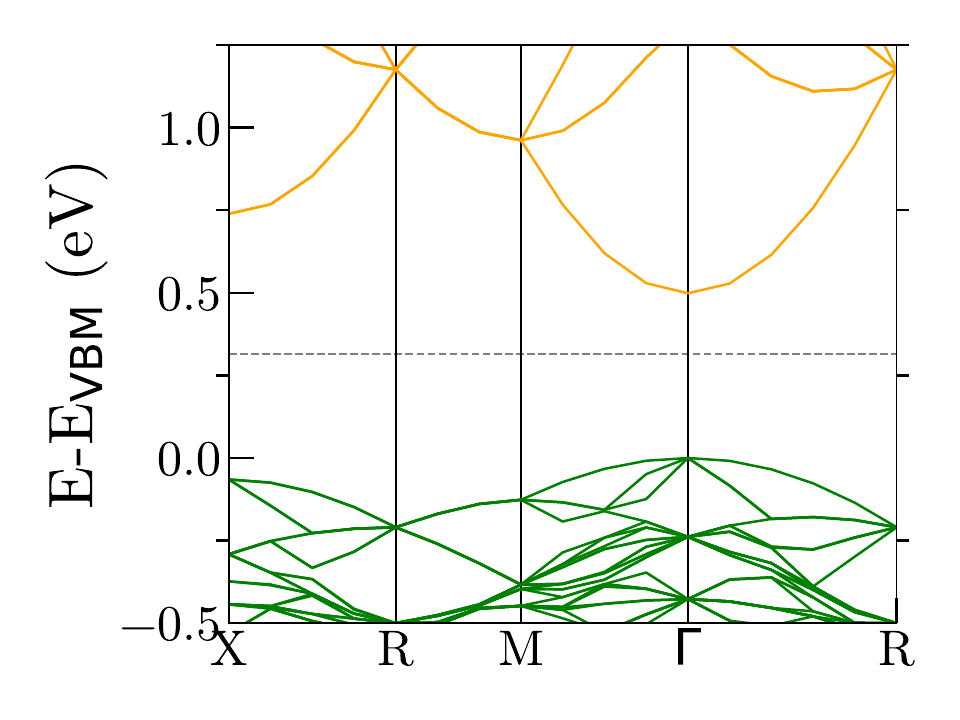}
    \includegraphics[width=0.9\linewidth]{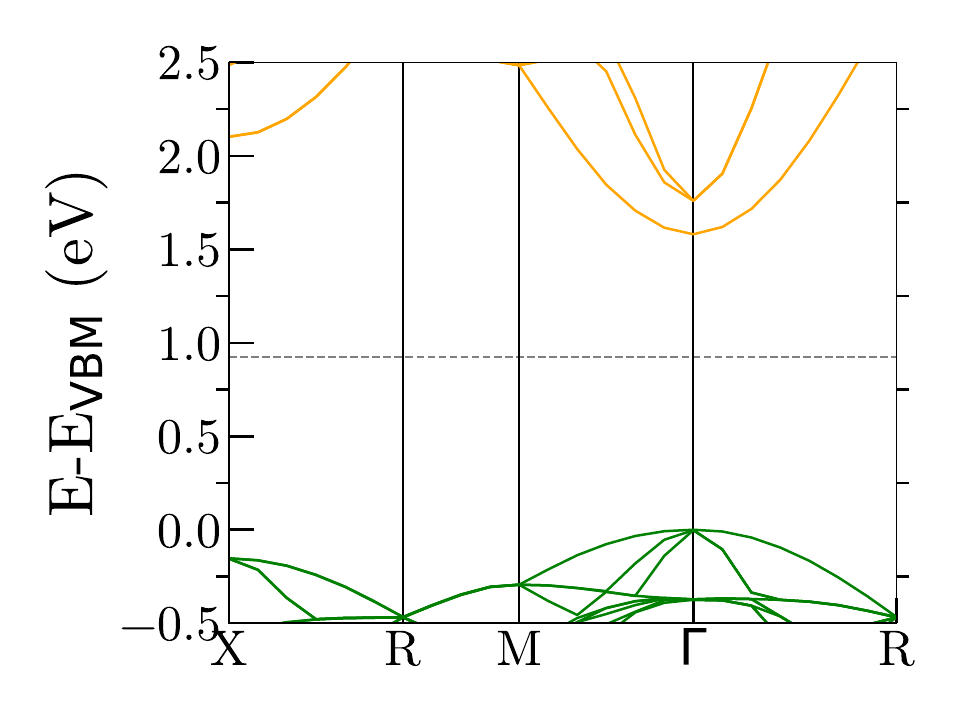}
    \caption{Band structures for pure Cu$_2$O in supercells. From top to bottom: PBE band structure in a $2\times 2\times 2$ supercell; PBE band structure in a $3\times 3\times 3$ supercell; HSE06 band structure in a $2\times 2\times 2$ supercell. Occupied states are shown in green and unoccupied states are shown in orange. The dotted line denotes the Fermi level.}
    \label{fig:cu2o_supercells}
\end{figure}

In figure \ref{fig:cu2o_supercells} we calculate the band structures of the pure crystal for the $2\times2\times2$ and $3\times3\times3$ supercells under PBE, and the $2\times2\times2$ supercell under HSE06 without spin-orbit coupling. Band structures are a way of mapping the energy-momentum space of the electrons, and are constructed by plotting the energy eigenvalues along a path around the Brillouin zone. The path taken in this work is between  the high symmetry points $X=(1,0,0)\pi/a$, $R=(1,1,1)\pi/a$, $M=(1,1,0)\pi/a$, and $\Gamma=(0,0,0)$, where $a$ is the lattice parameter of the respective supercell. The focus here is on the gamma point, since that is where the gap in Cu$_2$O is located.

As expected for semilocal XC functionals, the PBE gap (0.50 eV) underestimates the experimental value (2.18 eV), underscoring the importance of cross-checking defect levels with HSE06 where feasible, but in good agreement with other PBE studies of Cu$_2$O~\cite{raebiger2007origins,soon2009}. Inaccurate band gaps are a problem inherent in Kohn-Sham DFT, but such a significant underestimation is an indication that Cu$_2$O is a strongly-correlated material, hence why it is important to include hybrid functionals like HSE06 in our analysis. Despite this limitation, PBE-DFT is known to reproduce well the shapes and orderings of bands, and so these are the elements that we extract from our simulations. 

The band gap calculated under HSE is $E_\text{g}^\text{HSE} = 1.60$~eV, which is closer to the experimental gap but as expected still not exact. Our result is less than the HSE gap found by Scanlon~\cite{scanlon2009} (2.12~eV), but the discrepancy is accounted for by the different HSE parameters used in that study. The comparison between PBE and HSE06 band structures also sets the energy scale against which defect levels must be judged. In particular, only features that persist under both functionals are considered credible in-gap states; features that shift into the valence band or conduction band when switching from PBE to HSE06 are discarded as functional artefacts. Analysis of the density of states reveals that the top of the valence band is predominantly of copper 3d and oxygen 2p character, and the bottom of the conduction band is of copper 4s and copper 3d character. The interplay between the 4s and 3d orbitals is characteristic of strongly correlated materials, and the ground state occupancy of these orbitals is sensitive to our treatment of correlation effects, highlighting the importance of employing hybrid functionals.

It is important to note that the boundaries of the (first) Brillouin zone are at $\pm \pi/a$, so for larger supercells, the Brillouin zone is smaller. This has two effects on the band structure of larger supercells: first, bands appear flatter, because the x-axis covers less reciprocal-space distance. Second, since the bands `reflect' off the Brillouin zone boundaries, the reduced boundaries cause additional bands to appear in the band structure (known as band folding). Furthermore, the additional electrons present in the larger supercell adds additional bands as well. These effects can be seen by comparing the $2\times2\times2$ and $3\times3\times3$ supercell PBE band structures in figure \ref{fig:cu2o_supercells}. For these reasons, it is imperative that we compare each defect supercell with the corresponding pure supercell, in order to discern which effects arise only from the defect.

\section{The Defects}

In this section we present the defect-induced changes to the electronic structure. 
Because the observed PL spectrum depends sensitively on the presence of mid-gap states, the identification of which defects do or do not produce such states is essential for interpreting sample-to-sample variability in Rydberg-exciton experiments.
We consider all the native single-point defects, plus the split vacancies i.e., the copper and oxygen vacancies, $V_\text{Cu}$ and $V_\text{O}$, where a single atom is missing from the lattice; the anti-sites, Cu$_\text{O}$ (O$_\text{Cu}$) where a copper (oxygen) atom is in the usual place of an oxygen (copper) atom; the interstitial atoms, Cu$^\text{oct}_\text{i}$, Cu$^\text{tet}_\text{i}$, O$^\text{oct}_\text{i}$ and O$^\text{tet}_\text{i}$, where a copper or oxygen atom is inserted into the stoichiometric lattice in one of two different sites, labelled tetrahedral and octahedral after the way they coordinate with neighbouring copper atoms; and the split copper vacancies, $V_\text{Cu}^\text{s,1}$ and $V_\text{Cu}^\text{s,2}$, defect complexes formed from two adjacent copper vacancies with a copper interstitial at their midpoint. $V_\text{Cu}^\text{s,1}$ is formed when the copper vacancies are formed around different oxygen atoms, and $V_\text{Cu}^\text{s,2}$ is formed when the copper vacancies are formed around the same oxygen atom. Only $V_\text{Cu}^\text{s,1}$ has been previously studied using DFT~\cite{scanlon2009,soon2009,isseroff2013,dongfang2023}.

Based on the criteria we set out in section \ref{sec:criteria}, we sort the results from each defect into 3 catagories: no gap states, gap states, and non-robust states. The defects in the first category display no new states between the VBM and the CBM regardless of the choice of supercell size or functional (PBE/HSE). The defects in the second category generate additional low dispersion, well localised states within the pure band gap at all levels of theory. The third category is for defects that do not fully meet the criteria, for example exhibiting gap states only under the HSE functional, or gap states whose dispersion is too high to infer the behaviour in the dilute limit. Within each category we consider each defect in detail.

\subsection{Defects with no new states in the band gap}
Four defects did not exhibit any additional states within the pure Cu$_2$O band gap under any of the conditions we tested, including both PBE and HSE06 functionals and both $2\times2\times2$ and $3\times3\times3$ supercells. For these defects, the band structures of the defective and pristine supercells were indistinguishable within the robustness criteria. No dispersionless bands appeared between the bulk VBM and CBM and no features persisted upon supercell enlargement or functional change. These defects therefore cannot be responsible for any of the experimentally observed sub-band-gap photoluminescence lines.
\begin{center}
    \centering
    \begin{tabular}{ c c c c c }
        & V$_\text{Cu}$ & V$_\text{O}$ & V$_\text{Cu}^{s;2}$ & O$_\text{Cu}$ \\[0.5ex]
        \hline
        $\Delta H_\text{F}$, eV (Cu rich) & 0.35 & 0.66 & 1.35 & 3.75 \\
        $\Delta H_\text{F}$, eV (Cu poor) & -0.08 & 1.16 & 1.02 & 2.92 \\
        Charge & -1 & 0 & -1 & -1 \\
        PBE, $2\times2\times2$ & No & No & No & No \\
        PBE, $3\times3\times3$ & No & No & No & No \\
        HSE06, $2\times2\times2$ & No & No & No & No \\
    \end{tabular}
    \captionof{table}{Summary of defects with no gap states. We provide the formation enthalpy and charge of the lowest energy state, and indicate whether (Yes) or not (No) each level of theory predicts defect states in the band gap.}
    \label{tab:hf_no}
\end{center}

\subsubsection{Copper vacancy}

\begin{figure}
    \centering
    \includegraphics[width=0.9\linewidth]{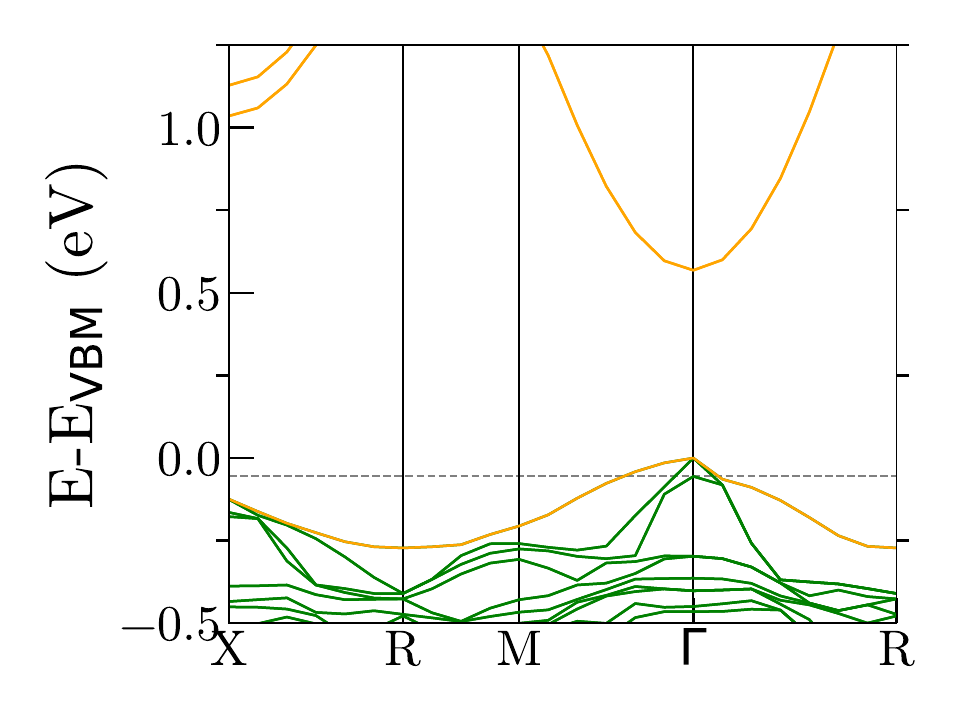}
    \includegraphics[width=0.9\linewidth]{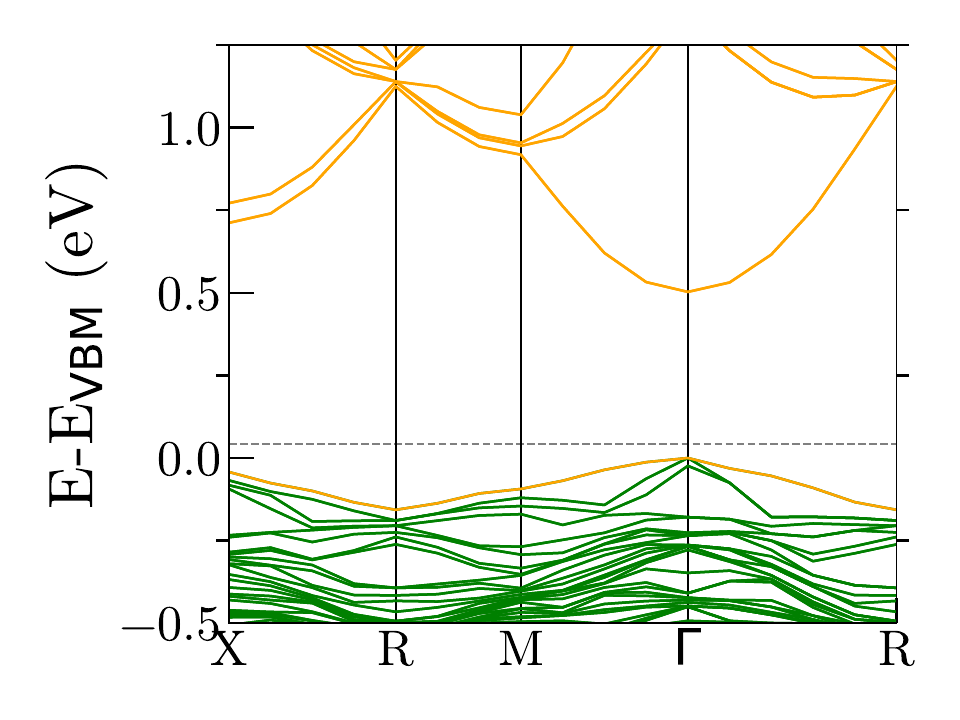}
    \includegraphics[width=0.9\linewidth]{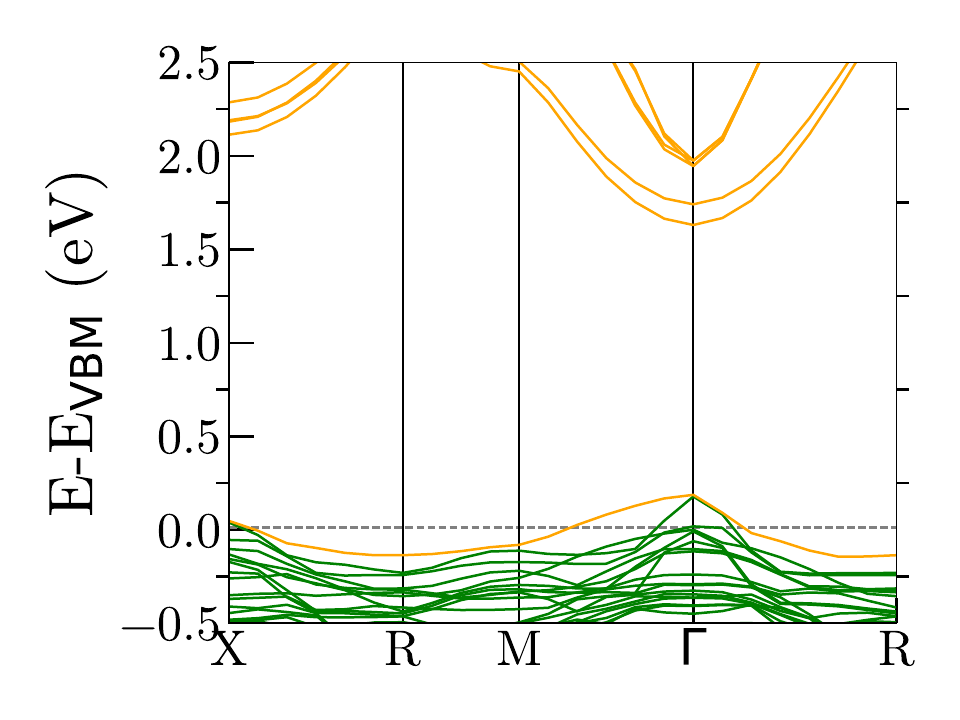}
    \caption{Band structures for the simple copper vacancy, $V_\text{Cu}$. From top to bottom: PBE band structure in a $2\times 2\times 2$ supercell; PBE band structure in a $3\times 3\times 3$ supercell; HSE06 band structure in a $2\times 2\times 2$ supercell. Occupied bulk states are shown in green and unoccupied bulk states are shown in orange. The dotted line denotes the Fermi level. The simple copper vacancy does not exhibit defect states in the band gap.}
    \label{fig:hse_vcu}
\end{figure}

The band structures for the simple copper vacancy for $2\times 2\times 2$ and $3\times 3\times 3$ supercells under PBE and for the $2\times 2\times 2$ supercell under HSE06 are shown in figure \ref{fig:hse_vcu}. The occupied states are shown in green and the unoccupied states are shown in orange. We see that $V_\text{Cu}$ does not introduce a clear electronic state in the band gap in any of the calculations. In the smaller supercell we see that $V_\text{Cu}$ perturbs the valence band such that the VBM becomes slightly peaked under both PBE and HSE06. Were we only to look at the eigenvalues at the $\Gamma$ point, it would be easy to believe that a defect state has appeared in the band gap just above the valence band. However, when we dilute the defect further in the $3\times 3\times 3$ supercell, we observe the structure return to be identical to the valence and conduction bands in the bulk crystal (fig. \ref{fig:cu2o_supercells}), demonstrating that the peaking of the valence band is merely a finite-size effect caused by an unphysical effective density of defects. Given the similarity between the PBE and HSE upper valence bands in the $2\times 2\times 2$ supercell, we can be confident that HSE is not introducing new physics which is absent from the PBE level of analysis, and that the peak in the valence band under HSE is also a finite-size effect. 

Across all three calculations we also see that the top of the valence band is partially unoccupied. This is expected, since the removal of a copper atom leaves an odd number of electrons. Under PBE the VBM is spin degenerate and half occupied, but under HSE the top state is actually only of one spin channel, a consequence of the spin splitting caused by Hartree-Fock exchange near the Fermi level. Neither effect would produce photoluminescence lines of energy less than $E_g$, but, as has been reported many times before~\cite{scanlon2009,raebiger2007origins,nolan2006}, these defects are responsible for much of the p-type conductivity observed in the semiconductor literature on cuprous oxide~\cite{al2015review}. 

For the PBE $3\times3\times3$ supercell, charges of up to $\pm2$ electrons had no effect on the band structure of V$_\text{Cu}$, other than to occupy or un-occupy the bands seen in the neutral band structures. Therefore, there is no reason to believe charged V$_\text{Cu}$ would produce any PL lines either. These results provide strong evidence that V$_\text{Cu}$ does not produce any mid-gap states under any realistic charge state, supercell size, or XC-functional, and therefore cannot be the origin of the PL peak at 1.35~eV.

\subsubsection{Oxygen vacancy}

\begin{figure}
    \centering
    \includegraphics[width=0.9\linewidth]{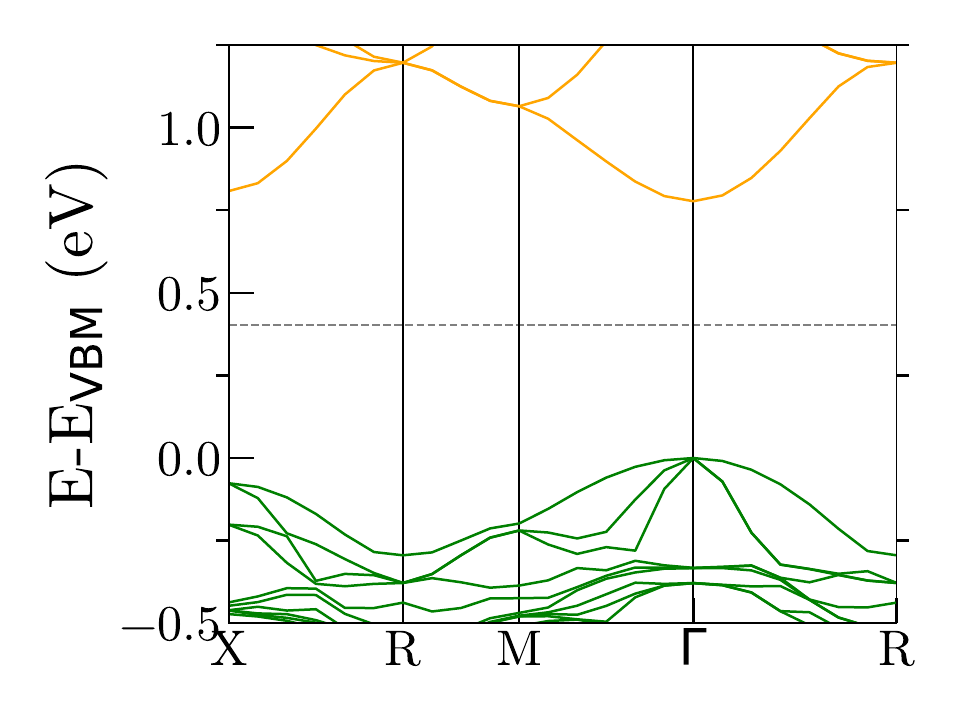}
    \includegraphics[width=0.9\linewidth]{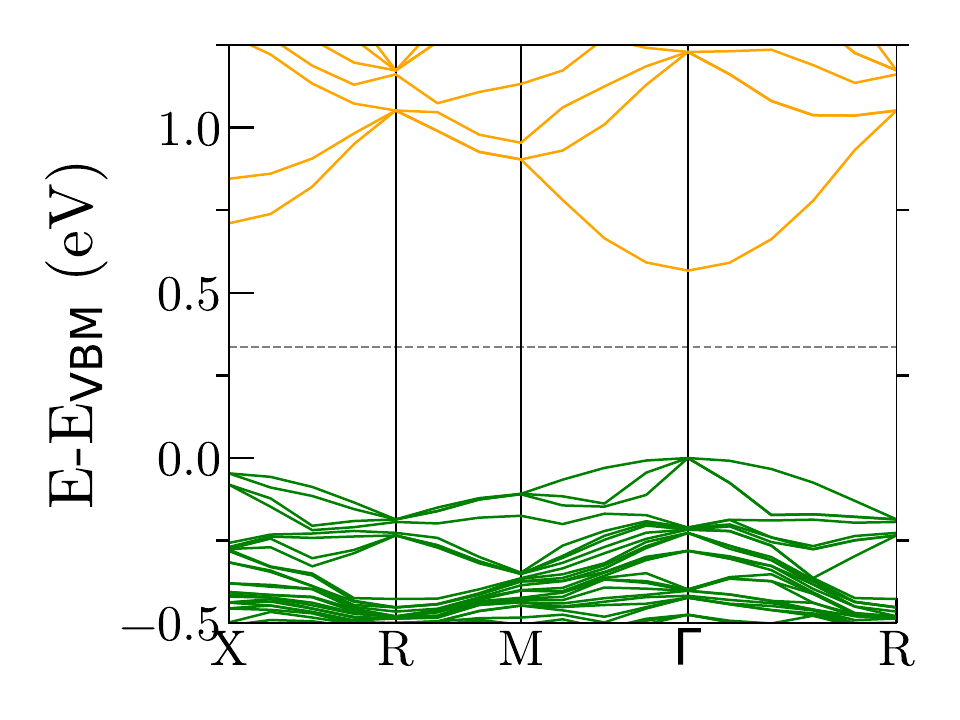}
    \includegraphics[width=0.9\linewidth]{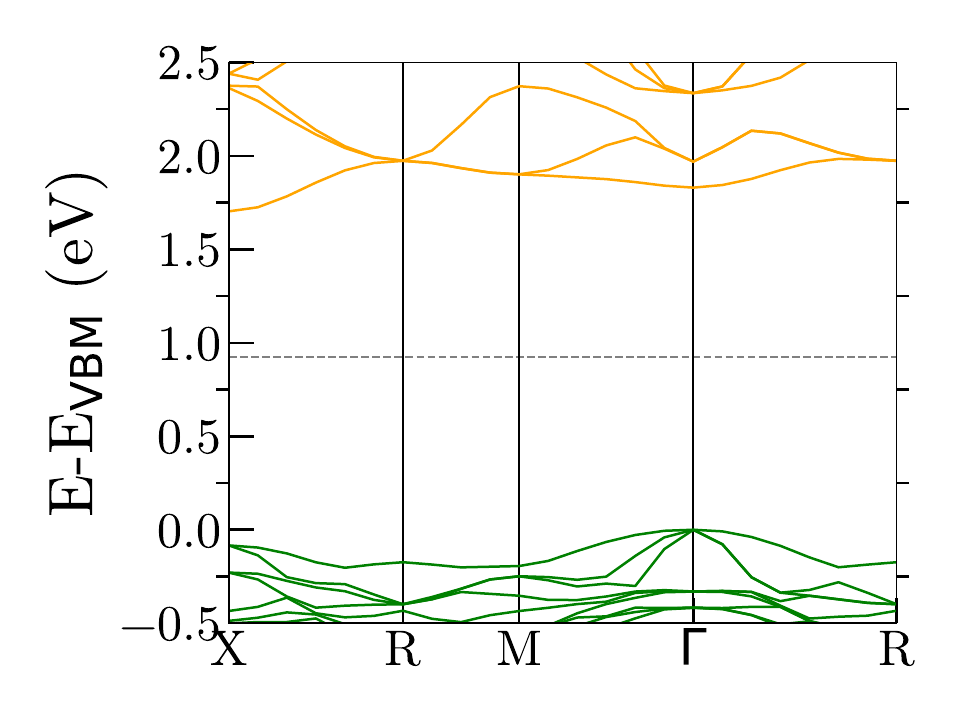}
    \caption{Band structures for the oxygen vacancy, $V_\text{O}$. From top to bottom: PBE band structure in a $2\times 2\times 2$ supercell; PBE band structure in a $3\times 3\times 3$ supercell; HSE06 band structure in a $2\times 2\times 2$ supercell. Occupied bulk states are shown in green and unoccupied bulk states are shown in orange. The dotted line denotes the Fermi level. The oxygen vacancy does not exhibit defect states in the band gap.}
    \label{fig:hse_vo}
\end{figure}

As seen in figure \ref{fig:hse_vo}, the oxygen vacancy also produces no gap state across all 3 levels of analysis. In the $2\times2\times2$ supercells, the effect of the oxygen vacancy (when compared to the pure crystal) is to flatten the CBM and open the gap, to $0.78$~eV (PBE) and $1.83$~eV (HSE). However, the PBE $3\times3\times3$ supercell band structure returns almost exactly to the bulk behaviour, with a gap of $0.56$~eV, demonstrating that the effects seen in the smaller supercells are finite-size effects, not genuine defect states. While it is hard to be certain how dilute the defect would need to be for the gap to return to its bulk value, we can be certain that larger supercells will not introduce states inside the band gap if they are not seen in these artificially high defect concentrations.

For the PBE $3\times3\times3$ supercell, charges of up to $\pm2$ electrons had no effect on the band structure of V$_\text{O}$, other than to occupy or empty the bands observed in the neutral defect. Since no gap state appears in the neutral defect supercell, there is no gap state associated with the charge oxygen vacancy either. Therefore, there is no reason to believe charged V$_\text{O}$, including V$_\text{O}^{+}$ and V$_\text{O}^{2+}$, would produce any PL lines either.



\subsubsection{Other}
The second form of the split copper vacancy, $V^\text{s,2}_\text{Cu}$, as well as the oxygen anti-site, O$_\text{Cu}$, also produce no defects states in the band gap across all 3 levels of analysis. In fact they both behave very similarly to $V_\text{Cu}$, including the peaking of the VBM in small supercells and introducing holes into the valence band. Charges of $\pm2$e also had no material effect on their band structures. The band structures for these defects can be seen in figures \ref{fig:ocu} and \ref{fig:vcus2} in the appendix.



\subsection{Defect with additional states in the band gap}
Two defects gave rise to clear additional state(s) in the band gap at all levels of analysis:
\begin{center}
    \centering
    \begin{tabular}{ c c c }
        & O$_\text{i}^\text{tet}$ & O$_\text{i}^\text{oct}$ \\[0.5ex]
        \hline
        $\Delta H_\text{F}$, eV (Cu rich) & 1.56 & 2.00 \\
        $\Delta H_\text{F}$, eV (Cu poor) & 1.06 & 1.50 \\
        Charge & +1 & 0 \\
        PBE, $2\times2\times2$ & Yes & Yes \\
        PBE, $3\times3\times3$ & Yes & Yes \\
        HSE06, $2\times2\times2$ & Yes & Yes \\
    \end{tabular}
    \captionof{table}{Summary of defects that show a state in the band gap. We provide the formation enthalpy and charge of the lowest energy state, and indicate whether (Yes) or not (No) each level of theory predicts defect states in the band gap.}
    \label{tab:hf_yes}
\end{center}

\subsubsection{Oxygen interstitials}

\begin{figure}
    \centering
    \includegraphics[width=0.9\linewidth]{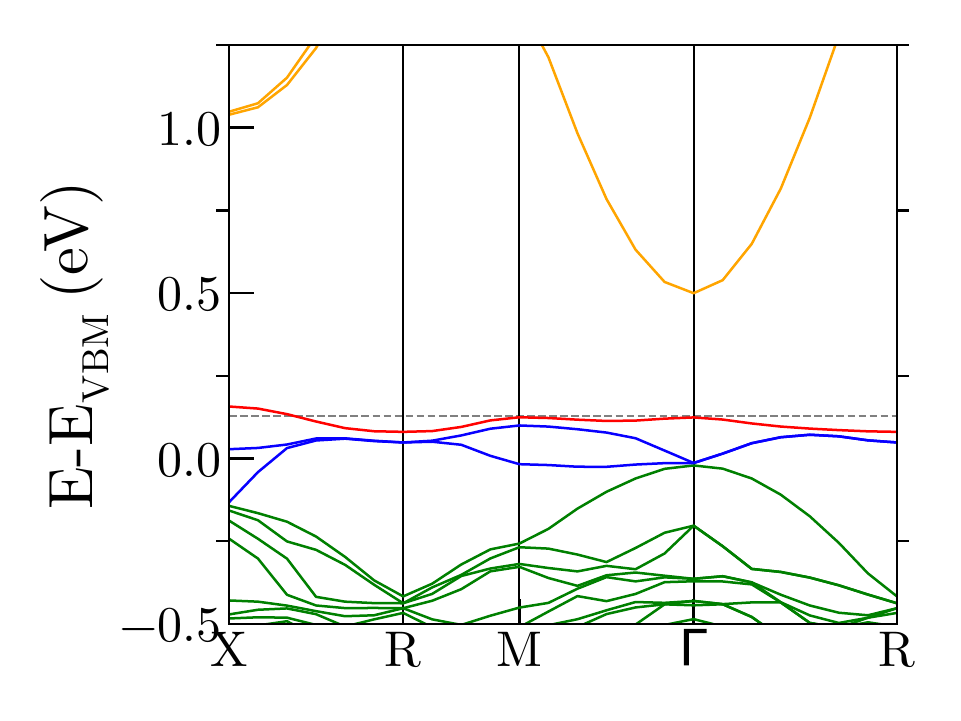}
    \includegraphics[width=0.9\linewidth]{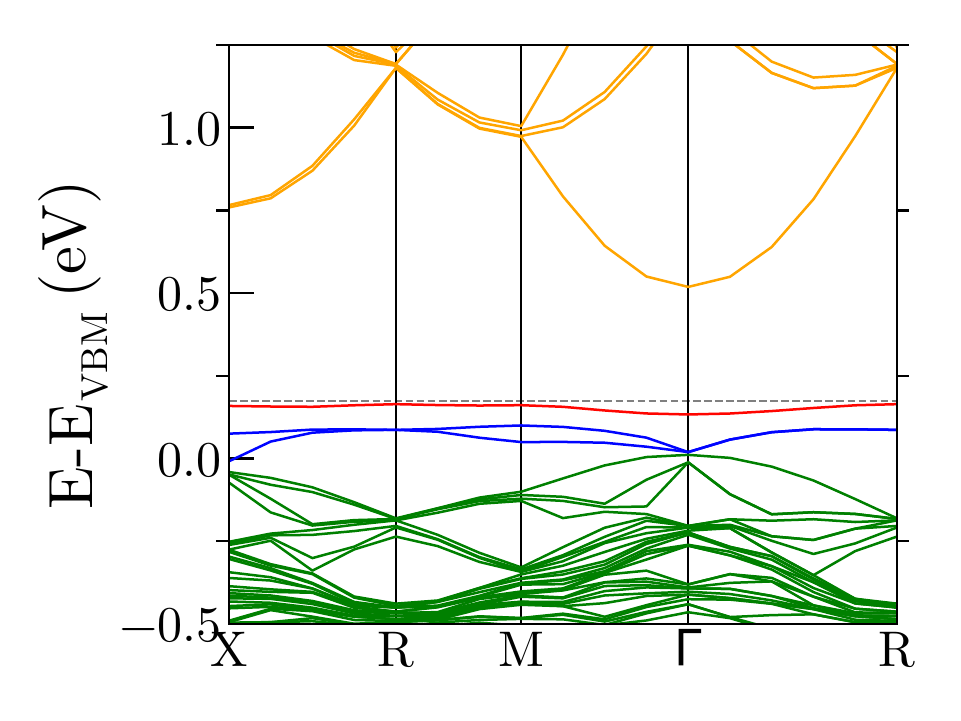}
    \includegraphics[width=0.9\linewidth]{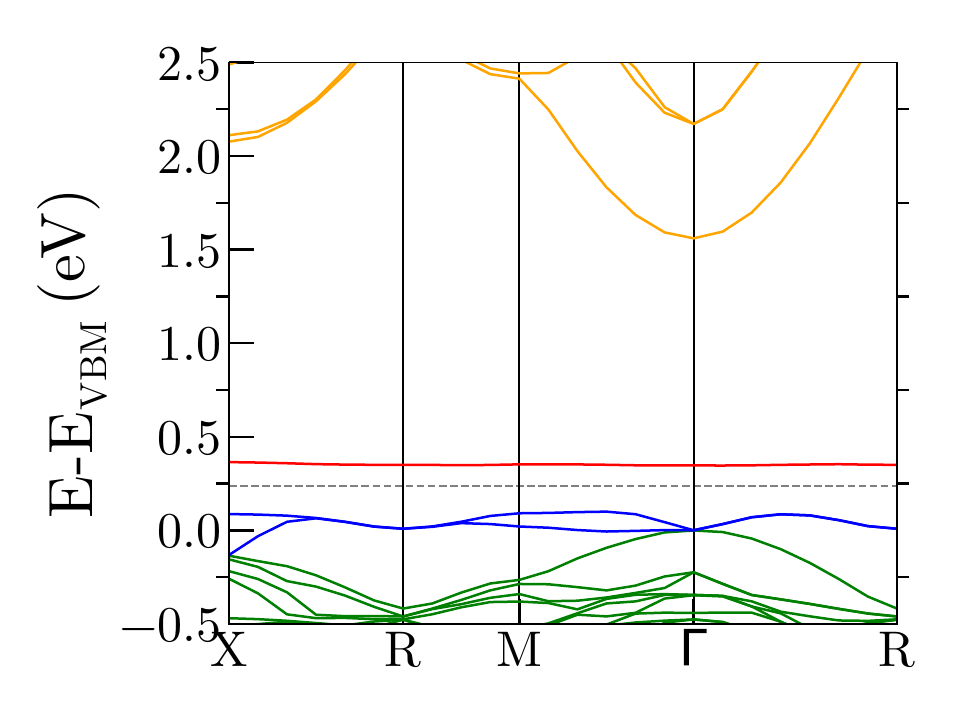}
    \caption{Band structures for the octahedral form of the oxygen interstitial, O$_\text{i}^\text{oct}$. From top to bottom: PBE band structure in a $2\times 2\times 2$ supercell; PBE band structure in a $3\times 3\times 3$ supercell; HSE06 band structure in a $2\times 2\times 2$ supercell. Occupied bulk states are shown in green, unoccupied bulk states in orange, occupied defect states in blue and unoccupied defect states in red.}
    \label{fig:hse_oi_oct}
\end{figure}

\begin{figure}
    \centering
    \includegraphics[width=0.9\linewidth]{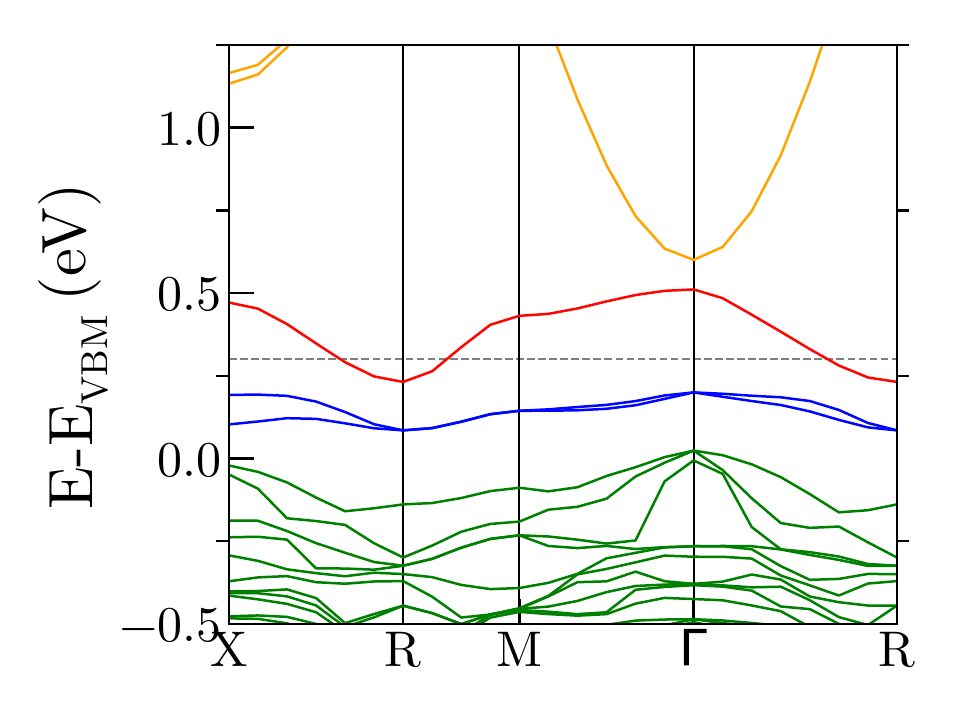}
    \includegraphics[width=0.9\linewidth]{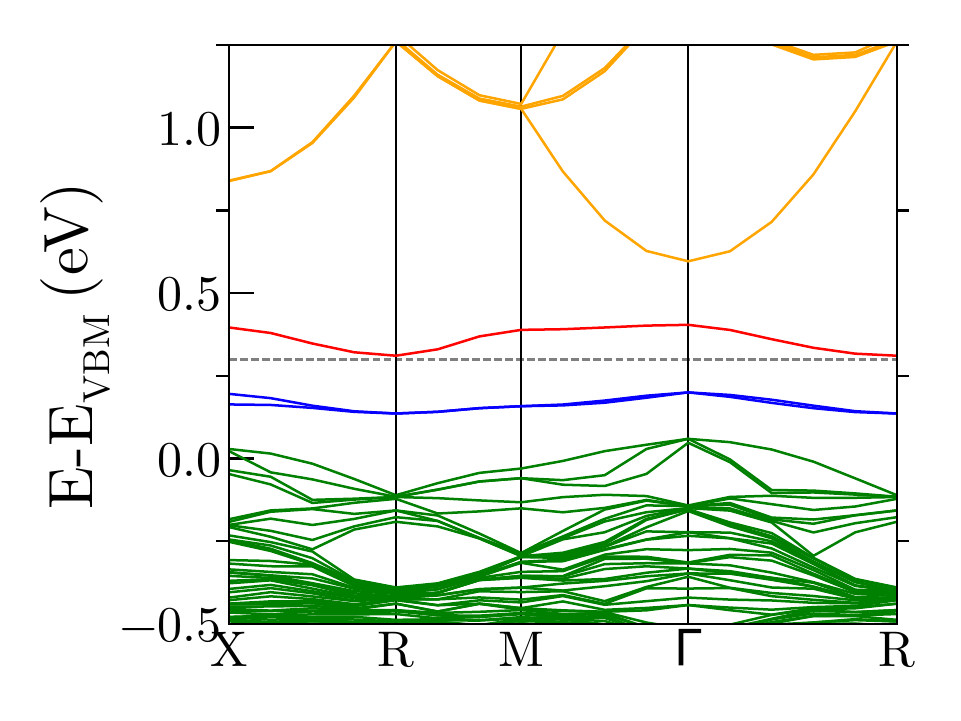}
    \includegraphics[width=0.9\linewidth]{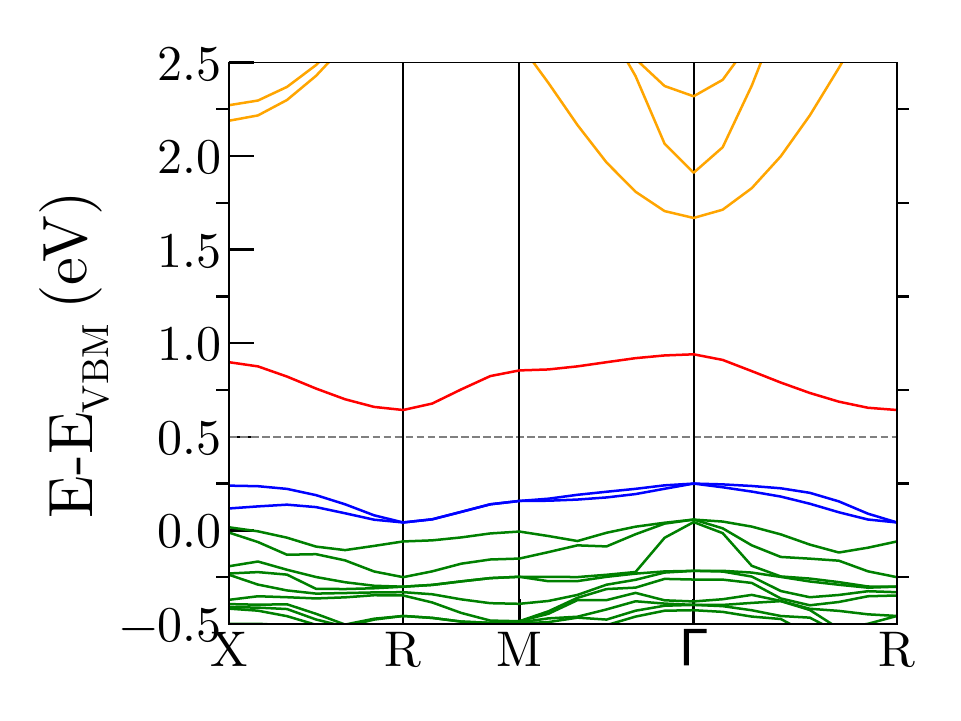}
    \caption{Band structures for the tetrahedral form of the oxygen interstitial, O$_\text{i}^\text{tet}$. From top to bottom: PBE band structure in a $2\times 2\times 2$ supercell; PBE band structure in a $3\times 3\times 3$ supercell; HSE06 band structure in a $2\times 2\times 2$ supercell. Occupied bulk states are shown in green, unoccupied bulk states in orange, occupied defect states in blue and unoccupied defect states in red.}
    \label{fig:hse_oi_tet}
\end{figure}

Both interstitial geometries show clear defect states in the band gap across both supercell sizes and for both PBE and HSE XC-functionals. The corresponding band structures are shown in figures \ref{fig:hse_oi_oct} and \ref{fig:hse_oi_tet}. Both geometries produce three well localised spin-degenerate states in the gap, 2 occupied and 1 unoccupied, which are easily understood as the 2P electrons of the additional oxygen atom.

As can be seen for O$_\text{i}^\text{oct}$ (fig. \ref{fig:hse_oi_oct}), the unoccupied defect state remains deep in the band gap for the $3\times 3\times 3$ supercell. Along with the fact that it is so flat, meaning it is very well localised, we can be confident it is not a consequence of neighbouring defects interacting with each other but a genuine, additional, local electronic state that would exist in the dilute defect limit. The two occupied defect states for O$_\text{i}^\text{oct}$ are degenerate with the VBM at $\Gamma$. Across all three calculations the defect had no effect on the band gap, lending further evidence that the states would not move in relation to the band gap in the dilute limit. 

In contrast, figure \ref{fig:hse_oi_tet} shows that all three defect states introduced by O$_\text{i}^\text{tet}$ exist firmly inside the appropriate bulk gap and could even produce several distinct photoluminescence lines in experiments. All three states localise further from the $2\times 2\times 2$ supercell to the $3\times 3\times 3$ supercell. The gaps in the $2\times 2\times 2$ supercells are slightly larger than their bulk values, but in the $3\times 3\times 3$ supercell the band gap returns to the bulk PBE value; this effect is likely to also occur for HSE.

\begin{figure}
    \centering
    \includegraphics[width=\linewidth]{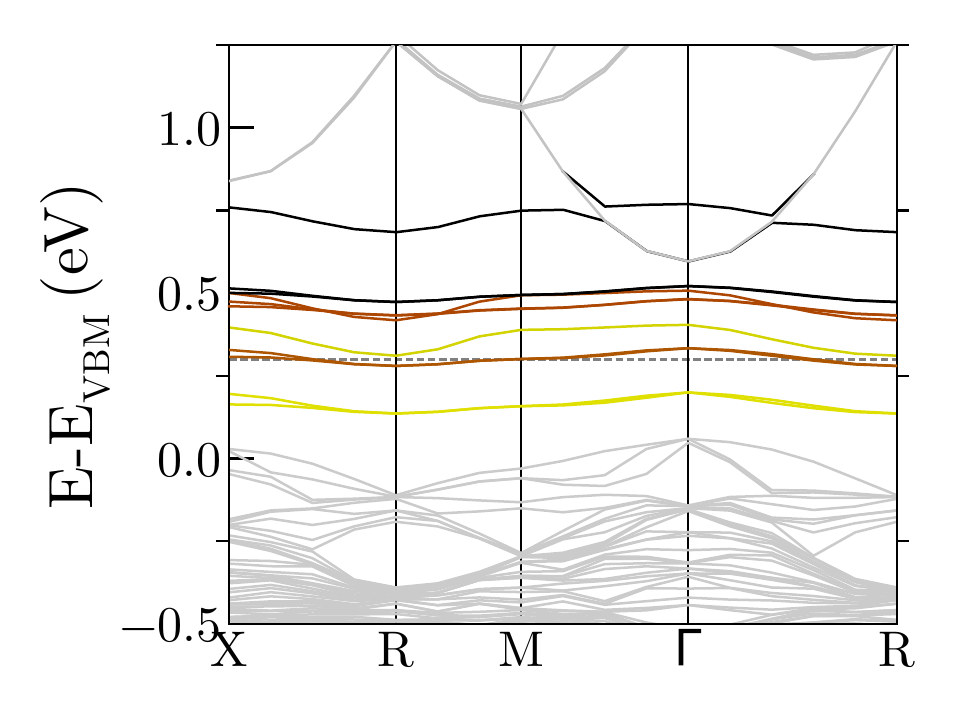}
    \caption{The effect of charge on the PBE $3\times3\times3$ supercell band structure of the tetrahedral oxygen interstitial, O$_\text{i}^\text{tet}$. The defect states in yellow are of the neutral defect, brown for $\big($O$_\text{i}^\text{tet}\big)^{-}$, and black for $\big($O$_\text{i}^\text{tet}\big)^{2-}$.}
    \label{fig:oi_t_charge}
\end{figure}

For the oxygen interstitials in the $3\times3\times3$ PBE supercell, we found that all 3 defect levels shift rigidly up and down for each electron added or removed respectively for  up to $\pm2$ electrons. This indicates that the different charge states of these defects could give rise to distinct lines in the PL spectrum. The shift was $\sim 0.2$~eV for O$_\text{i}^\text{tet}$ and $\sim 0.1$~eV for O$_\text{i}^\text{oct}$ per electron. The exception to this rule was (O$_\text{i}^\text{oct}$)$^{+}$, which maintained the same energy of in-gap defect state as O$_\text{i}^\text{oct}$. The effect on the O$_\text{i}^\text{tet}$ band structure is shown in figure \ref{fig:oi_t_charge}.

\subsection{Non-robust defect states}
Another group of defects produced features that \emph{appeared} within or near the band gap under at least one level of theory, but failed to satisfy the robustness criteria. In these cases, apparent in-gap signatures were strongly dependent on the exchange–correlation functional, the supercell size, or both, and typically vanished when moving to the larger $3\times3\times3$ cell or when switching from PBE to HSE06. Because these features cannot be distinguished from band-folding artefacts or finite-size distortions, we classify these defects as non-robust and do not consider them viable candidates for any of the observed PL lines. In each instance, the putative defect state merged into the valence band, shifted above the CBM or developed strong dispersion upon refinement indicating that it was not a localized defect-induced level. Their properties are summarised here:
\begin{center}
    \centering
    \begin{tabular}{ c c c c c }
        & Cu$_\text{i}^\text{tet}$ & Cu$_\text{i}^\text{oct}$ & V$_\text{Cu}^\text{s;1}$ & Cu$_\text{O}$ \\[0.5ex]
        \hline
        $\Delta H_\text{F}$, eV (Cu rich) & 2.17 & 1.36 & 0.62 & 3.80 \\
        $\Delta H_\text{F}$, eV (Cu poor) & 2.50 & 1.69 & 0.28 & 4.63 \\
        Charge & +1 & +1 & -1 & +1 \\
        PBE, $2\times2\times2$ & Yes & Yes & No & No\\
        PBE, $3\times3\times3$ & Non. & Non. & No & No\\
        HSE06, $2\times2\times2$ & Yes & Yes & Yes & Non.\\
    \end{tabular}
    \captionof{table}{Summary of non-robust defects: We provide the formation enthalpy and charge of the lowest energy state, and indicate whether (Yes) or not (No) each level of theory predicts defect states in the band gap. ``Non.'' is short for non-robust.}
    \label{tab:hf_maybe}
\end{center}

\subsubsection{Copper interstitials}
\begin{figure}
    \centering
    \includegraphics[width=0.9\linewidth]{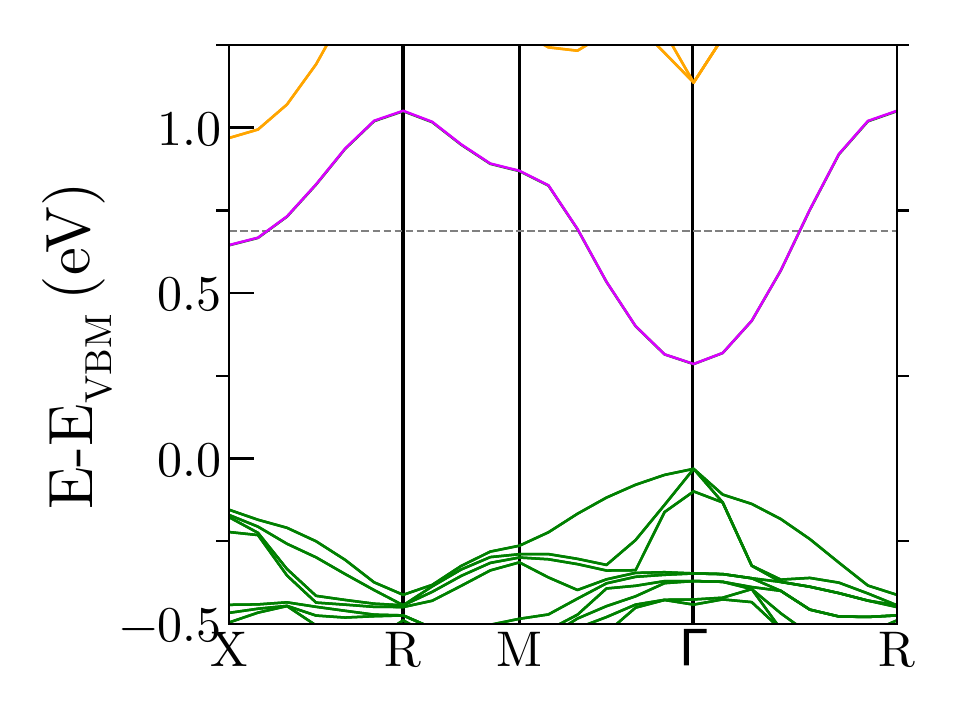}
    \includegraphics[width=0.9\linewidth]{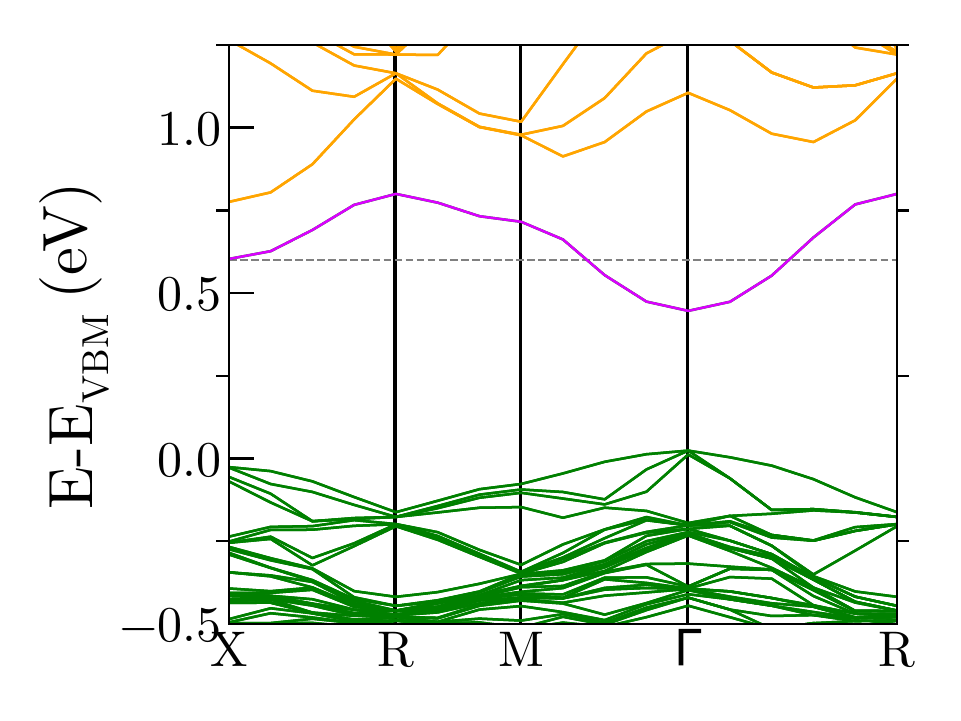}
    \includegraphics[width=0.9\linewidth]{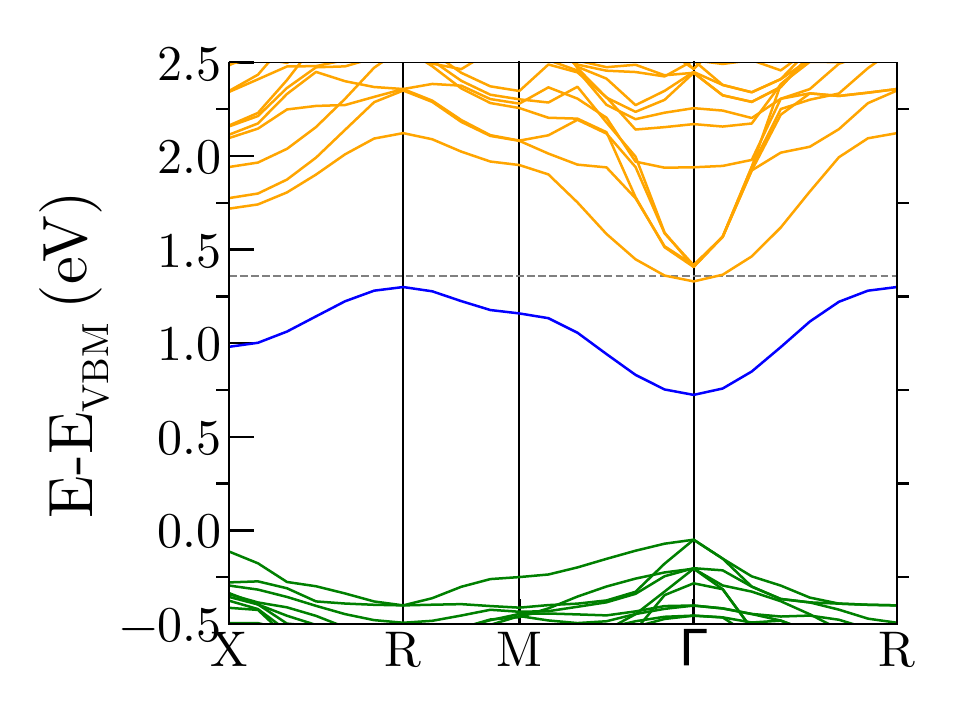}
    \caption{Band structures for the octahedral form of the copper interstitial, Cu$_\text{i}^\text{oct}$. From top to bottom: PBE band structure in a $2\times 2\times 2$ supercell; PBE band structure in a $3\times 3\times 3$ supercell; HSE06 band structure in a $2\times 2\times 2$ supercell. Occupied bulk states are shown in green, unoccupied bulk states in orange, occupied defect states in blue and half-occupied defect states in purple.}
    \label{fig:hse_cui_oct}
\end{figure}

\begin{figure}
    \centering
    \includegraphics[width=0.9\linewidth]{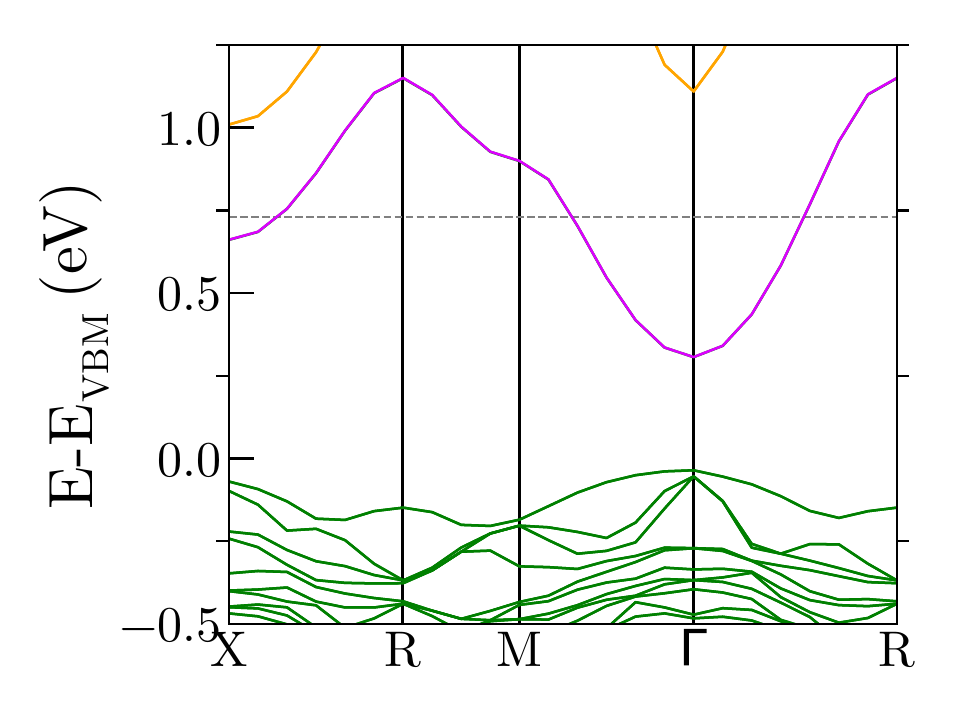}
    \includegraphics[width=0.9\linewidth]{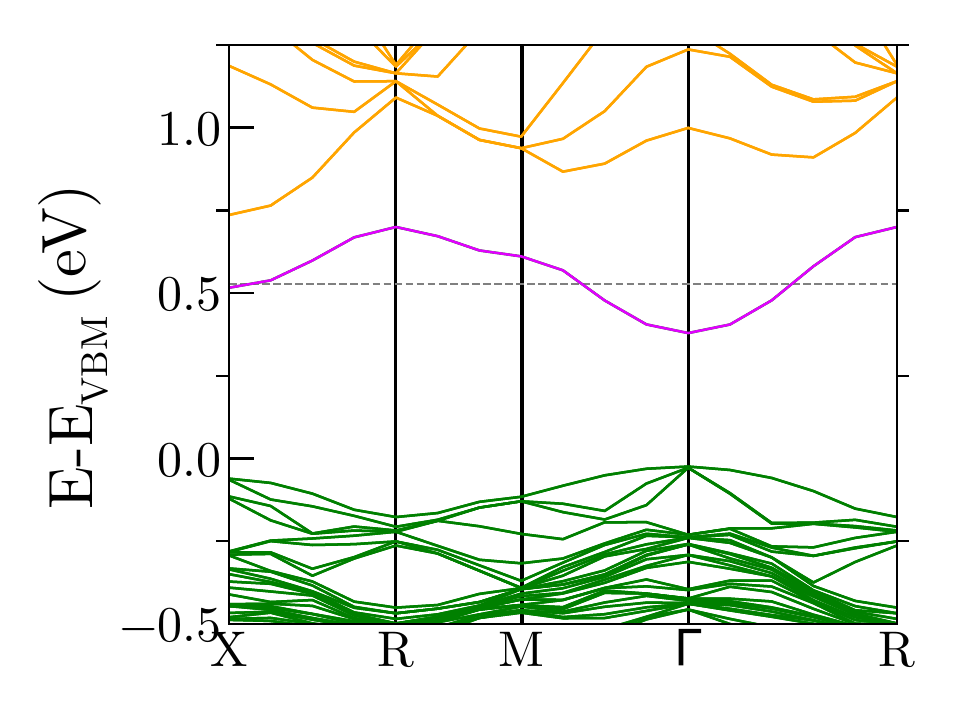}
    \includegraphics[width=0.9\linewidth]{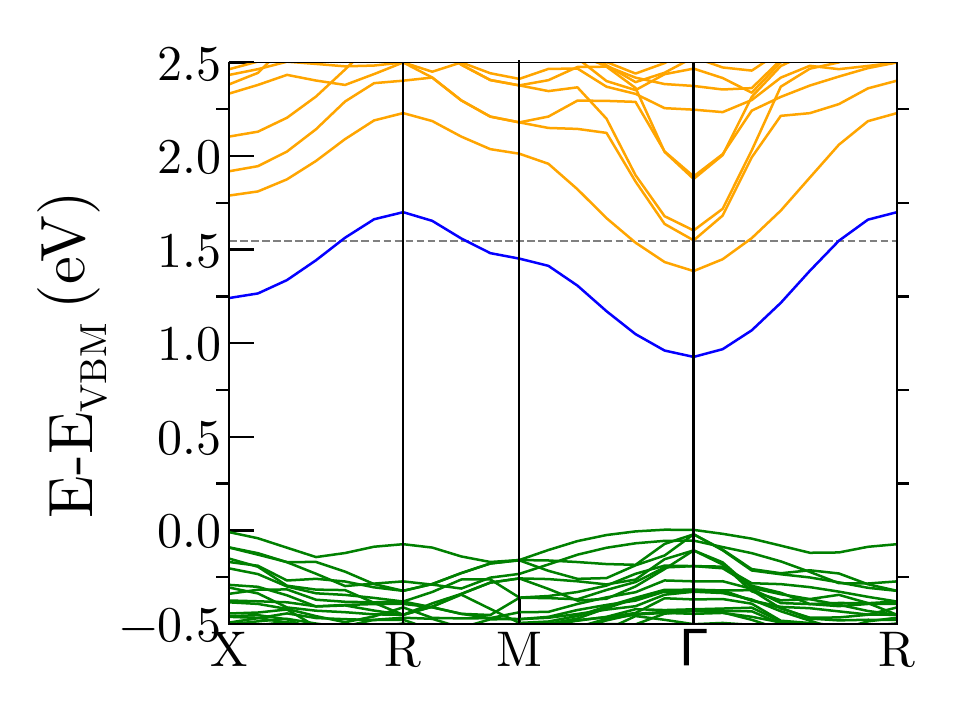}
    \caption{Band structures for the tetrahedral form of the copper interstitial, Cu$_\text{i}^\text{tet}$. From top to bottom: PBE band structure in a $2\times 2\times 2$ supercell; PBE band structure in a $3\times 3\times 3$ supercell; HSE06 band structure in a $2\times 2\times 2$ supercell. Occupied bulk states are shown in green, unoccupied bulk states in orange, occupied defect states in blue and half-occupied defect states in purple.}
    \label{fig:hse_cui_tet}
\end{figure}

The copper interstitials, both in the octahedral (figure \ref{fig:hse_cui_oct}) and tetrahedral (figure \ref{fig:hse_cui_tet}) arrangements, produce similar distortions from the bulk structure, so we will discuss them together. 

Under PBE, for the $2\times 2\times 2$ supercell, a half-occupied state appears at $0.32$~eV above the VBM for the octahedral arrangement and $0.34$~eV for the tetrahedral arrangement, well within the bulk PBE gap at $\Gamma$. The state has a very wide dispersion, indicating a large extent in real space, and so a large finite-size effect. Comparing these band structures to the bulk crystal, we also see that the CBM is missing, so we could think that these defect states are simply a perturbation of the CBM that would return to bulk behaviour in the dilute limit.

Interestingly, the defect states are somewhat robust to increasing the supercell size, indicating that this state could exist even in the dilute defect limit. The band increases to $0.42$~eV for octahedral and $0.40$~eV for tetrahedral for the $3 \times 3 \times 3$ PBE supercell, still less than but closer to the bulk PBE gap. Comparing again to the bulk $3 \times 3 \times 3$ PBE supercell band structure, a band appears at the bottom of the conduction band which could be the bulk CBM returning to its bulk behaviour. One can imagine two scenarios for the dilute limit under PBE: the defect state rejoins the conduction band as the bulk CBM; or the state remains within the PBE gap and the bulk CBM emerges from the conduction band. A further study of the $4 \times 4 \times 4$ supercell would be needed to distinguish between these two.

The HSE analysis gives us a different angle to understand these defects. The half-occupied state predicted by PBE is split by Hartree-Fock exchange into a lower occupied spin state and an upper unoccupied spin state. The occupied spin states are $0.93$~eV and $0.98$~eV above the VBM for Cu$^i_\text{oct}$ and Cu$^i_\text{tet}$ respectively, and the unoccupied ones not only resemble the bulk CBM but are exactly the bulk HSE gap above the VBM. 

For both copper interstitials, in the PBE $3\times3\times3$ supercell, charging the defect by adding (subtracting) up to $2$ electrons raised (lowered) the energy of the defect state by $0.05$~eV per electron. This introduces the possibility that different charge states of the copper interstitials could produce a cluster of neighbouring PL lines. However, the behaviour of the charged defects may be significantly different under HSE, given the large spin splitting of the defect level.

The strong dispersion of these states indicates that the electron wavefunctions are delocalised across the supercell, interacting at long range with periodic copies of the defect site. These finite-size effects violate our criterion for a genuine defect state, as we cannot be sure of the behaviour in the dilute limit. Our cautious interpretation of such results is that the interstitial copper atoms provide an additional, local binding of one of the CBM states, leading to a local state in the band gap, near in energy to the CBM, which extends over several unit cells in real space. However, larger supercell calculations would be required to verify this analysis.

\subsubsection{First split copper vacancy}
\begin{figure}
   \centering
    \includegraphics[width=0.9\linewidth]{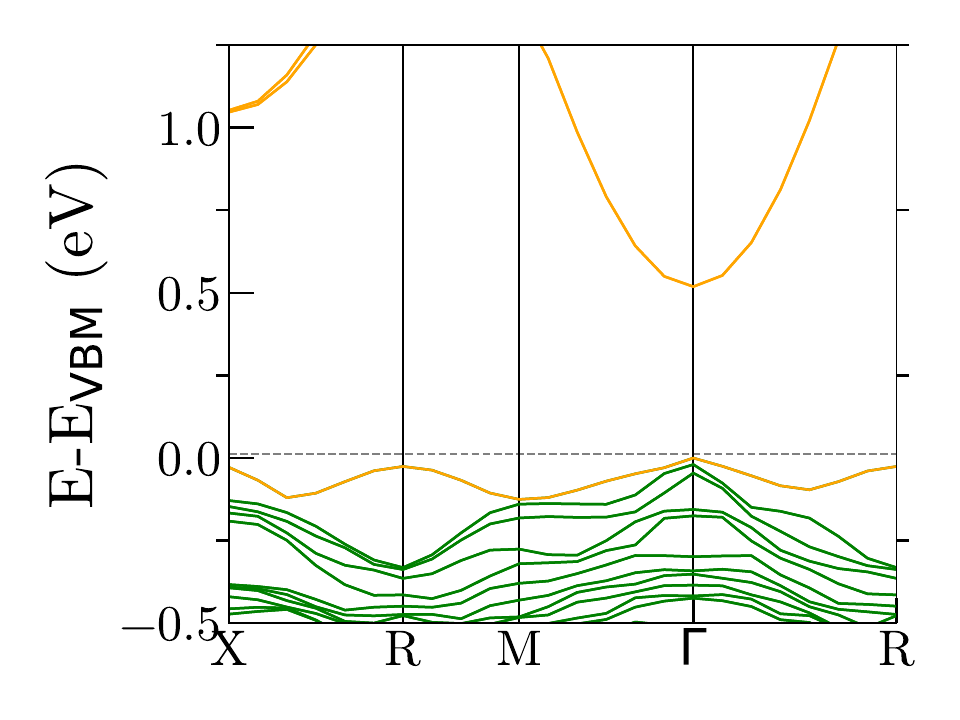}
    \includegraphics[width=0.9\linewidth]{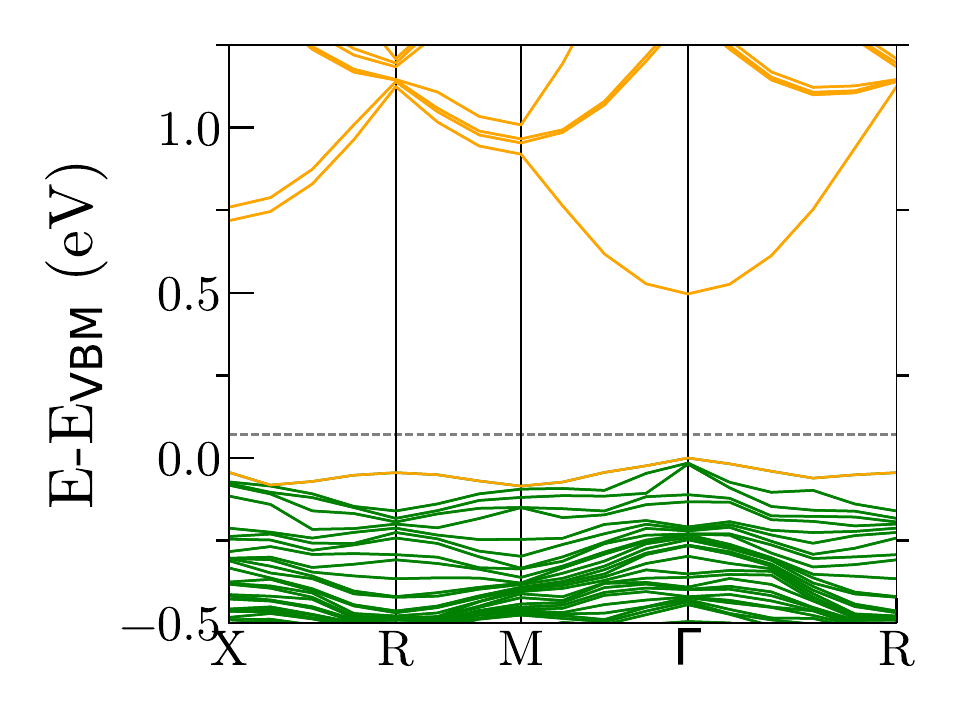}
   \includegraphics[width=0.9\linewidth]{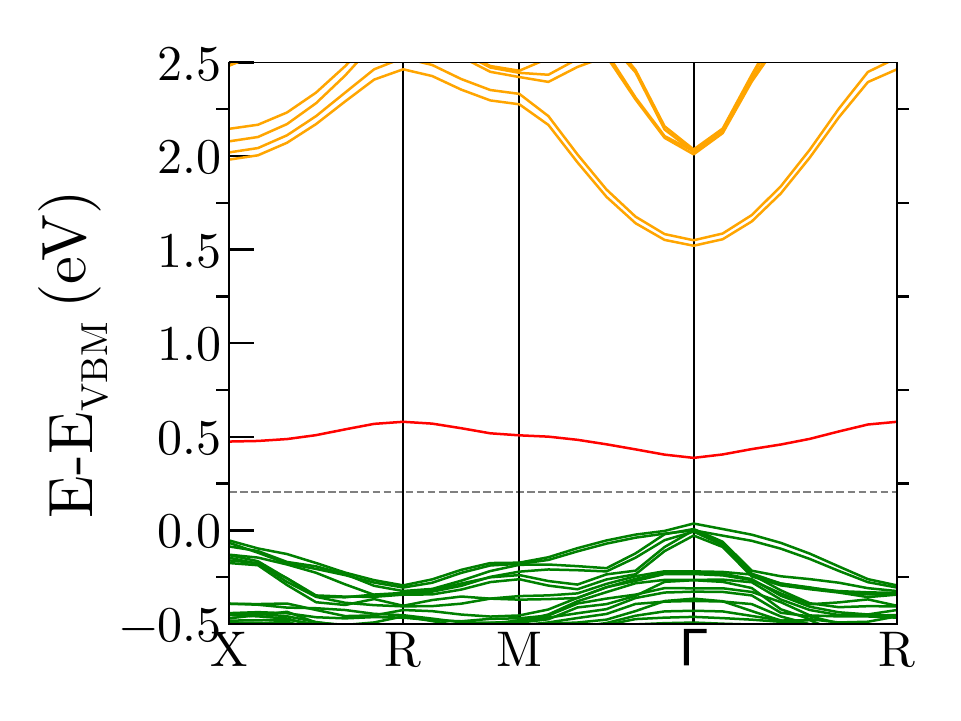}
    \caption{Band structures for the first form of the split copper vacancy, $V_\text{Cu}^\text{s,1}$. From top to bottom: PBE band structure in a $2\times 2\times 2$ supercell; PBE band structure in a $3\times 3\times 3$ supercell; HSE band structure in a $2\times 2\times 2$ supercell. Occupied bulk states are shown in green, unoccupied bulk states are shown in orange, and unoccupied defect states are shown in red. The dotted line denotes the Fermi level.}
    \label{fig:hse_vcus1}
\end{figure}

Under PBE, the band structure for the first split copper vacancy (where the copper atoms are bonded to different oxygen atoms), seen in figure \ref{fig:hse_vcus1}, looks very similar to the simple copper vacancy and the second split copper vacancy. The only difference is that the VBM for $V_\text{Cu}^{s;1}$ is raised at $R$ to the same energy as at $\Gamma$ in the $2\times 2\times 2$ supercell. The VBM mostly restored to that of the bulk in the $3\times 3\times 3$, but even if this persists in the dilute limit it will not have an effect on the photoluminescence. Like the other copper vacancies, the VBM also becomes partially unoccupied, contributing to native p-type conductivity, but not to photoluminescence.

Under HSE, a defect state appears in the HSE gap $0.35$~eV above the valence band maximum. The band is very flat and well localised, and only exists for one of the spin channels. This is in excellent agreement with Isseroff and Carter~\cite{isseroff2013}, who found a defect state with only one allowed spin with a line in the density of states $0.57$~eV above the VBM. The fact that this state only exists for one spin channel suggests a similar concern as for the copper interstitials. Hybrid functionals like HSE struggle to accurately treat partially-occupied states, since they encounter divergences in the derivative of the energy at the Fermi level, so this defect state could be an artefact. However, the other two copper vacancies have a partially-occupied VBM and do not exhibit such an artefact under HSE, which encourages us to believe the defect state for V$_\text{Cu}^{s;1}$ is physical and due to some physics captured by HSE not present under PBE.

Under PBE in the $3\times3\times3$ supercell, charges of up to $\pm2$ electrons had no effect on the band structure, other than to occupy or un-occupy the states present in the neutral band structure. Due to restrictions in computational resources, we did not investigate charge states under HSE, which for V$_\text{Cu}^{s;1}$ could be important. Since HSE can have problems treating states which cross the Fermi level, it is possible that the unoccupied defect level present under HSE in the neutral defect could return to the valence band when occupied. 

\subsubsection{Copper replacing oxygen}

The copper anti-site Cu$_\text{O}$ introduces a myriad of defect states in the gap, as seen in figure \ref{fig:cuo} in the appendix. Under PBE in the $2\times2\times2$ supercell, however, the gap increases to significantly above the bulk PBE gap, so we could not consider them potentially responsible for any PL lines. Under PBE in the $3\times3\times3$ supercell, the additional states flatten and localise, but remain at the energy of the CBM, so we do not expect below-gap PL in the dilute limit. Under HSE, there are many possible emitting states, but they display much of the same character as the states in the $2\times2\times2$ supercell under PBE, which changed dramatically when increasing to the $3\times3\times3$ supercell. Therefore, we do not assign any weight to the states seen under HSE. The effect of charge in the $3\times3\times3$ PBE supercell on Cu$_\text{O}$ is slightly complicated. Broadly, positive charge leaves no possible photoluminescent states, but negative charge does create a possible high-energy photoluminescent state. However, with formation energies of 4-5~eV, these states are very unlikely to manifest in the crystal.

\begin{table*}
\centering
\begin{tabular}{ c c c c c c c c c c c }
    & $V_\text{Cu}$ & $V_\text{Cu}^{s;1}$ & $V_\text{Cu}^{s;2}$ & $V_\text{O}$ & Cu$_\text{i}^\text{tet}$ & Cu$_\text{i}^\text{oct}$ & O$_\text{i}^\text{tet}$ & O$_\text{i}^\text{oct}$ & Cu$_\text{O}$ & O$_\text{Cu}$\\ [0.5ex] 
 \hline
    $\Delta H_\text{F}^{(q)}$, eV (O poor) & 0.24 & 0.62 & 1.35 & 0.66 & 2.17 & 1.36 & 1.56 & 2.00 & 3.80 & 3.75\\ [0.5ex]
    $\Delta H_\text{F}^{(q)}$, eV (O rich) & -0.08 & 0.29 & 1.01 & 1.16 & 2.50 & 1.69 & 1.06 & 1.50 & 4.63 & 2.92\\ [0.5ex]
    Charge & -1 & -1 & -1 & 0 & +1 & +1 & +1 & 0 & +1 & -1\\ [0.5ex]
    PBE \hspace{0.2cm} $2\times2\times2$ & No & No & No & No & Yes & Yes & Yes & Yes & No & No\\ [0.5ex]
    PBE \hspace{0.2cm} $3\times3\times3$ & No & No & No & No & Non. & Non. & Yes & Yes & No & No\\ [0.5ex]
    HSE06 $2\times2\times2$ & No & Yes & No & No & Yes & Yes & Yes & Yes & Non. & No\\ [0.5ex]
\end{tabular}
\caption{For each defect: the enthalpy of formation of the lowest enthalpy charge state at the two extremes of growth conditions; the charge of the lowest enthalpy charge state; whether or not the given level of theory predicts a defect state in the band gap. The abbreviation ``Non.'' is short for non-robust.}
\label{table:h_f}
\end{table*}



\section{Discussion}

There are 4 PL lines commonly observed across authors and samples~\cite{bloem1958, ito1997, frazer2015, frazer2017, koirala2013correlated, li2013engineering} at $\sim 1.2$~eV, $\sim 1.35$~eV, $\sim 1.5$~eV, and $\sim 1.7$~eV, and a fifth small line observed by Frazer et al~\cite{frazer2017} at $\sim 1.9$~eV. No one has yet identified clear trends in line sizes against growth conditions. Natural samples seem to be more often dominant in 1.7~eV emission~\cite{frazer2017, ito1997}, Li et al.~\cite{li2013engineering} and Koirala et al.~\cite{koirala2013correlated} synthesise samples that are dominant in 1.7~eV emission, and Frazer et al.~\cite{frazer2017}, Ito et al.~\cite{ito1998} and Bloem~\cite{bloem1958,biccariThesis} synthesise samples with all kinds of emissions. The emissions at 1.35~eV and the emission at 1.7~eV are usually stronger than the other 3, and the line at 1.35~eV is rarely seen without the shoulder at 1.2~eV. The only potential trend was documented by Bloem~\cite{bloem1958}, who found that the line at 1.7~eV disappeared in samples grown above a threshold oxygen pressure. However, with only 6 crystal samples, it is not a strong conclusion, and no other study investigates the relationship between PL and chemical growth conditions.

Conventionally, the line at 1.35~eV is assigned to the simple copper vacancy, V$_\text{Cu}$, and the lines at 1.5~eV and 1.7~eV are assigned to the oxygen vacancy, V$_\text{O}$. These assignments are due to Bloem~\cite{bloem1958}, and are based on 3 arguments: the only two meaningful native defects in Cu$_2$O are the simple copper and oxygen vacancies; Bloem observed the line heights at 1.5~eV and 1.7~eV to always be in the same ratio; and the peak at 1.7~eV disappeared at high oxygen pressure. The first two of these arguments are now know not to be true. As can be seen in table \ref{table:h_f} (and others studies~\cite{scanlon2009,soon2009}), many native defects can form in Cu$_2$O with formation enthalpies comparable to the vacancies, and therefore possibly contribute to PL. Furthermore, every other study of PL in Cu$_2$O~\cite{ito1998,frazer2015,frazer2017, koirala2013correlated, li2013engineering} has found the lines at 1.5~eV and 1.7~eV in a variety of different ratios. Therefore, the evidence for the assignment of V$_\text{Cu}$ and V$_\text{O}$ to these peaks is already weak.

Our state-of-the-art DFT calculations directly contradict this conventional assignment. There is no evidence from DFT that the oxygen vacancy $V_\text{O}$, simple copper vacancy $V_\text{Cu}$, second split copper vacancy $V_\text{Cu}^{s;2}$, or oxygen anti-site O$_\text{Cu}$ give rise to any states within the bulk band gap (table \ref{table:h_f}) nor would produce any signal in photoluminescence experiments. Therefore, there appears to be no reason to label the photoluminescence lines observed in experiments with either $V_\text{Cu}$ or with $V_\text{O}$. In fact, given their small perturbations in $2\times 2\times 2$ supercells for both XC-functionals and lack of effect on the band structure of the $3\times 3\times 3$ supercell, it is unlikely they would have any effect on excitons at all in the dilute limit. 

Our study provides strong evidence that the two geometries of the oxygen interstitial would result in one (octahedral) or more (tetrahedral) electronic states midway into the bulk band gap. Their low enthalpy of formation indicates they could produce strong emission lines in photoluminescence in defective samples grown in copper-poor conditions, which would be suppressed under copper-rich growth conditions. Their energies are also sensitive to the charge state of the defect, meaning they could be responsible for multiple emission lines. Further analysis of these defects with excited-state methods could be fruitful.

It is possible that the $V_\text{Cu}^{s;1}$ defect would produce a defect state in the band gap, and therefore a line in the photoluminescence, however the fact that the state only appears under HSE means it could be a non-physical artefact of the XC-functional. If it were physical, it's consistently very low formation enthalpy would make it a common defect, especially in samples grown in copper-poor conditions. On the other hand, the defect state only supports one electron per defect, as opposed to the two (spin up and spin down) electrons supported by each level from the oxygen interstitials, so the $V_\text{Cu}^{s;1}$ line may be suppressed.

It is also possible that the copper interstitials would produce defect states in the band gap. The effect of the defect is too long range to be properly converged in the $3\times 3\times 3$ PBE supercell, so it is difficult to say whether the half-occupied band is a perturbation of the CBM or a poorly localised defect state. The much stronger exchange treatment in HSE06 splits this band by spin and populates fully one of the spin channels, but hybrid functionals are known to do poorly with metal-like states such as this one~\cite{gao2016applicability}. Given how strongly these defects interact with neighbouring defects, it is also hard to say whether this splitting would decay to zero in the dilute limit. If the defect did have a state in the band gap, it is safe to say it would lie near the bottom of the conduction band and produce photoemission of energy similar to the band gap energy, $\sim 2$~eV. The much higher formation enthalpy of the tetrahedral site would indicate that the octahedral interstitial would be significantly brighter in PL.

Despite being well under-converged in supercell size in this study, the HSE results indicate it is possible that the copper anti-site could introduce a defect state into the band gap in the dilute limit. Nevertheless, its prohibitively high formation enthalpy would prevent it from contributing any signal to the photoluminescence spectrum.

\begin{figure}
    \centering
    \includegraphics[width=0.95\linewidth]{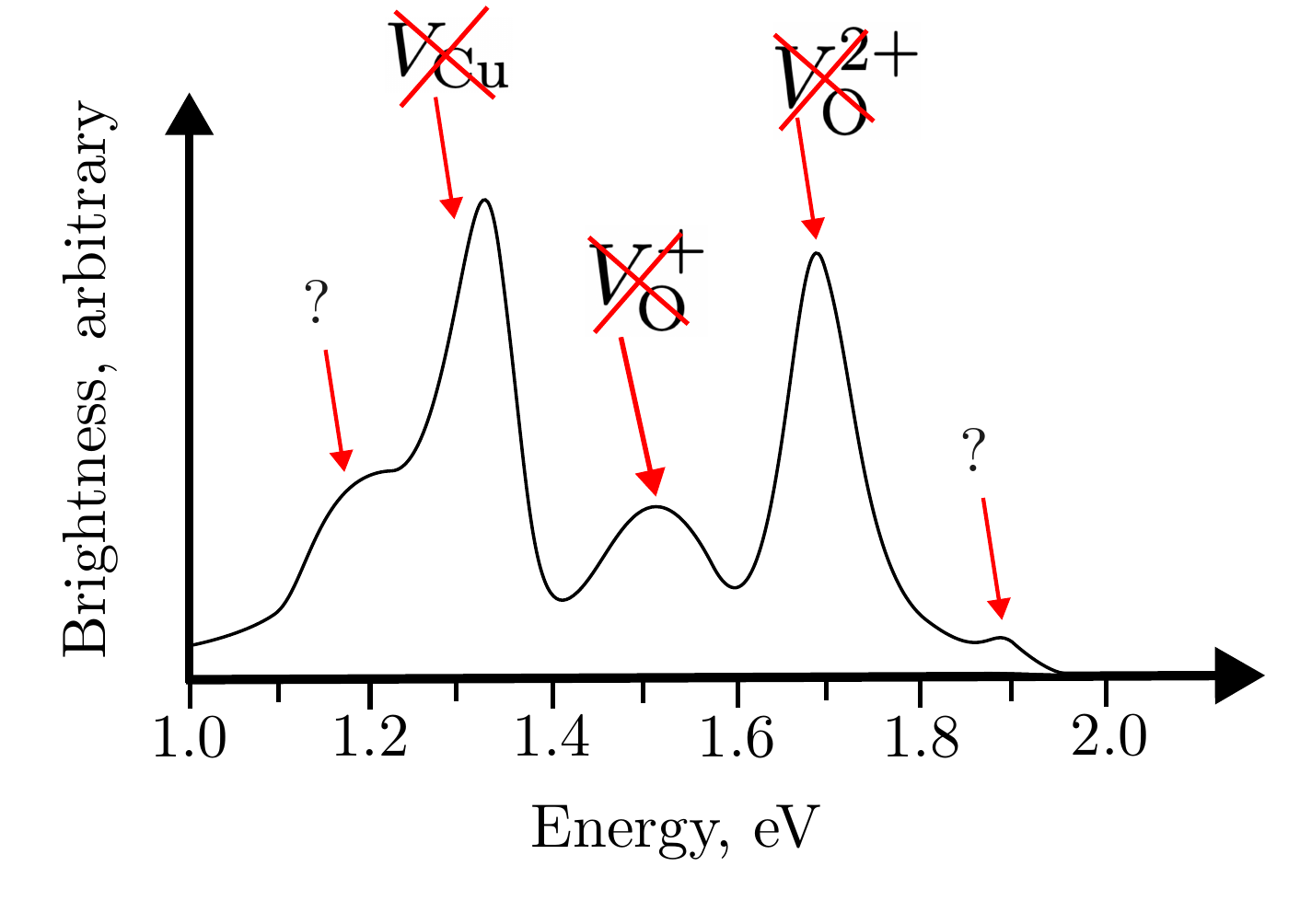}
    \includegraphics[width=0.95\linewidth]{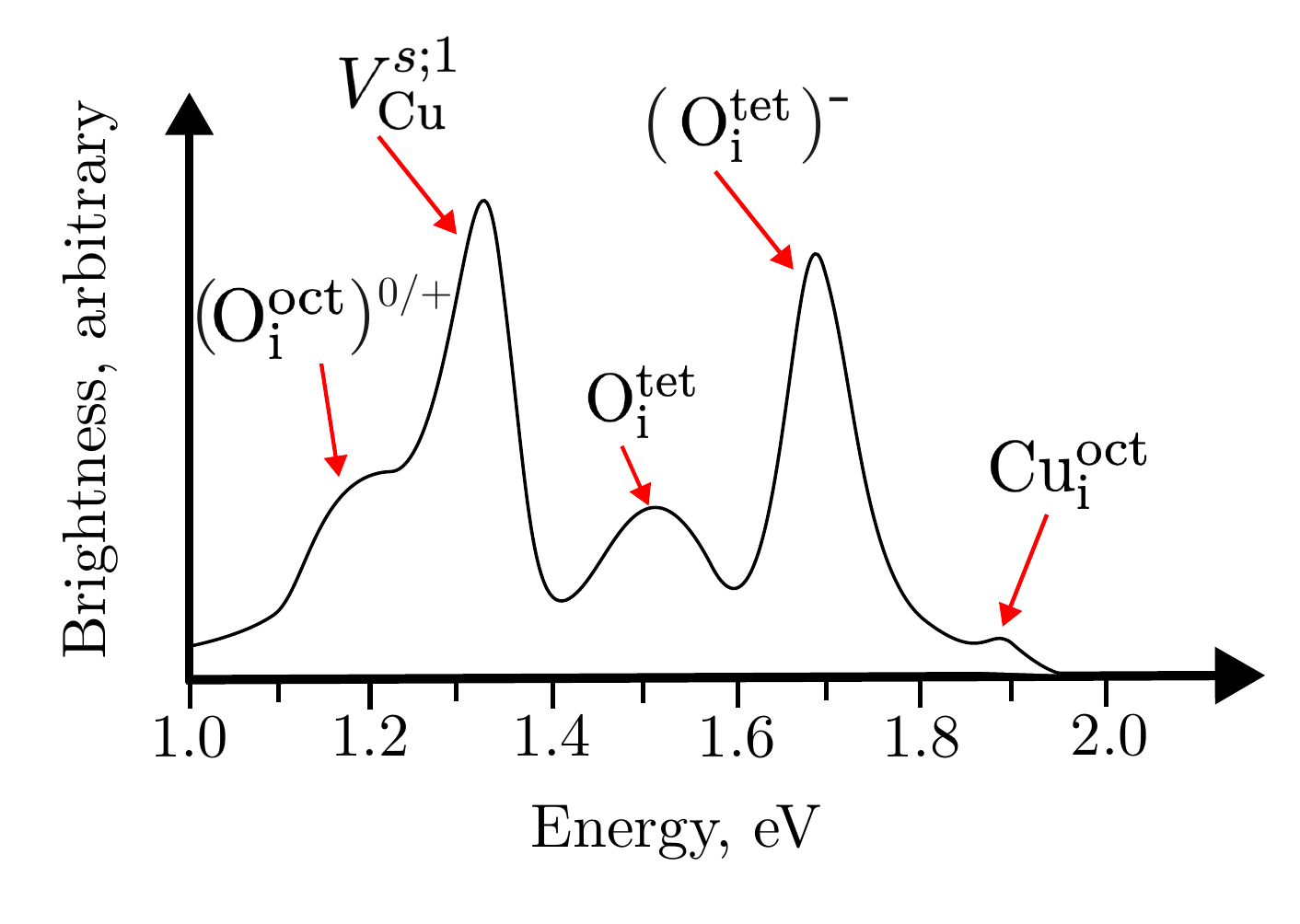}
    \includegraphics[width=0.95\linewidth]{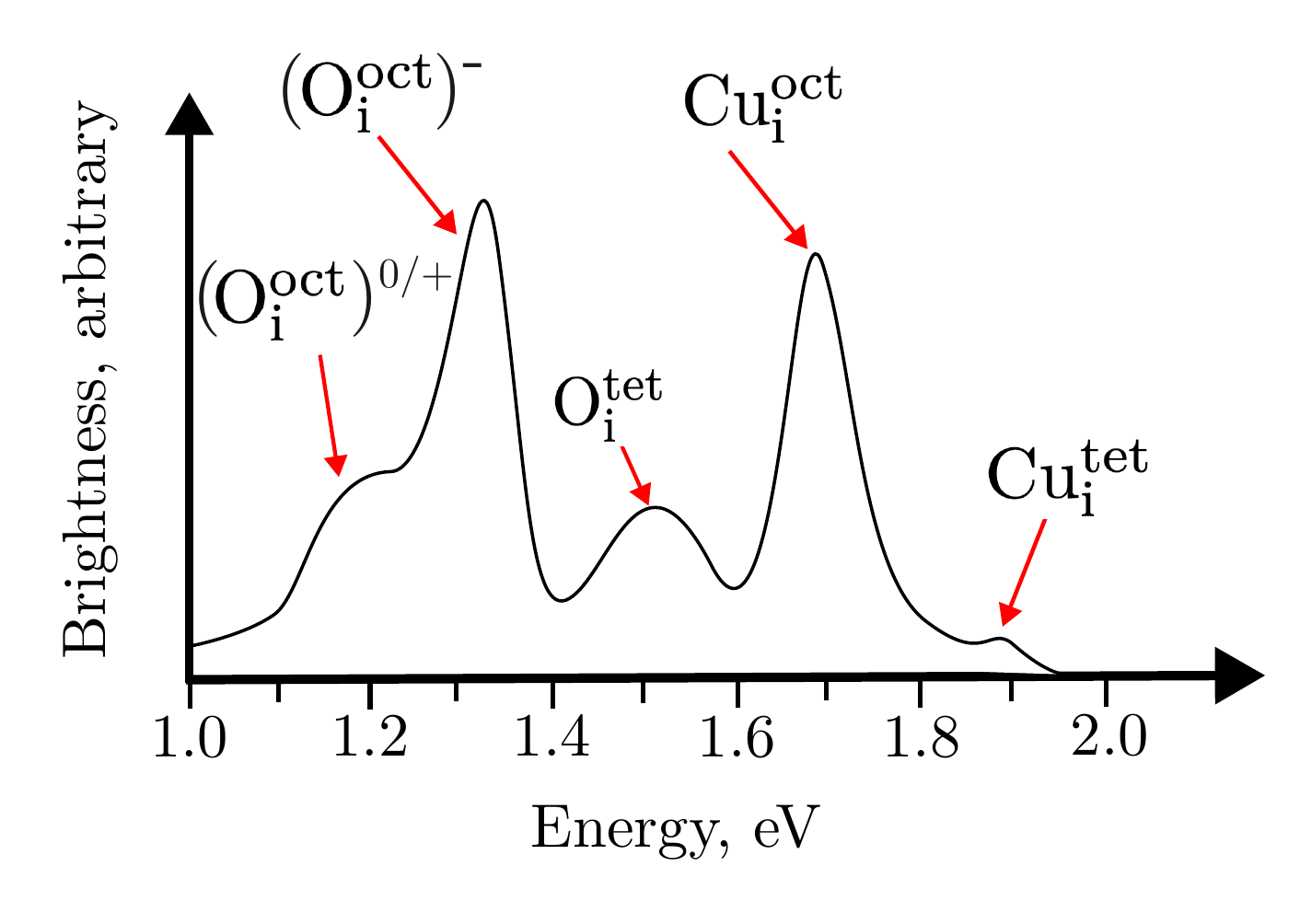}
    \caption{Sketches of all the commonly observed lines in photoluminescence experiments on Cu$_2$O. Top: the unsupported line assignments removed; middle: a possible reassignment of the PL lines based on the findings in this work; bottom: another possible assignment of the PL lines.}
    \label{fig:photoemission}
\end{figure}

Unfortunately, more than 5 possible photoluminescent states have emerged from this study, so we cannot conclude definitively on which PL lines are due to which defects. Six states come from each oxygen interstitial in charge states 0, -1, and -2; potentially two, close in energy, come from the copper interstitials; and potentially one comes from V$_\text{Cu}^{s;1}$. Given the trouble DFT has with accurately predicting band gaps, it is hard to directly relate the energy of the defect state to that of a photoemission line, especially given the additional error bar of $\pm 0.2$~eV from the lack of spin-orbit coupling. Even if we were to extend the PBE and HSE gap artificially to their experimental values, it is not clear whether we should keep the defect states at their gap above the VBM, at their gap below the CBM, or something else entirely. Having said that, PBE and HSE both seem to agree roughly on the fractional position of a given defect state within the respective gaps, so we use the approximate energy ordering of the states to suggest several plausible PL line assignments.

The lowest energies from this set are (O$_\text{i}^\text{oct})^{+/0/-}$ and V$_\text{Cu}^{s;1}$, so we are encouraged to assign the 1.2~eV and 1.35~eV lines to some of these states. These two lines also commonly seen together, which could be because V$_\text{Cu}^{s;1}$ and O$_\text{i}^\text{oct}$ both have a lower formation enthalpy under oxygen rich conditions, or because they are due to two or more charge states of O$_\text{i}^\text{oct}$. This second hypothesis is also supported by the fact that the two lines are separated by $\sim0.1$~eV, which is also how much the defect level rises by from (O$_\text{i}^\text{oct})^{0/+}$ to (O$_\text{i}^\text{oct})^{-}$. For the same reason, it is also possible that the line at 1.5~eV is due to (O$_\text{i}^\text{oct})^{2-}$. 

The next-lowest defect state energy is from O$_\text{i}^{tet}$, so it could be the cause of the 1.5 or 1.7~eV lines. Given the 0.2~eV rise in this defect level that results from adding electrons, it is also possible that the 1.7 and 1.9~eV lines are due to (O$_\text{i}^{tet})^{-}$ and (O$_\text{i}^{tet})^{2-}$. Under the hypothesis that all the lines are due to various charge states of the two geometries of oxygen interstitials, it is not clear what growth conditions could cause such varying concentrations of O$_\text{i}$ geometries and charges. The influence of growth conditions is easier to explain if we assign some of the high energy lines (1.7 and 1.9~eV) to the copper interstitials. Since, under PBE, they have very similar defect level energies, it is even possible that the two geometries contribute to the same PL line. 

Figure \ref{fig:photoemission} shows 2 possible peak re-assignments based on the evidence present in this study. As stated above, the results of this study are not sufficient to make concrete re-assignments of the PL lines to defect sites. Much stronger conclusions could be drawn from excited state calculations (GW or BSE) starting with HSE-DFT, as this would give quantitative emission spectra, with more accurate excitation energies attenuated by optical matrix elements.

For Rydberg-exciton experiments, this re-assignment is especially significant: interstitial defects generate stronger and more spatially localized potentials than vacancies, implying a different pattern of line broadening and impurity-induced Stark shifts than previously assumed.

Taken together, our results overturn the longstanding vacancy-based interpretation of the Cu$_2$O PL spectrum and instead identify interstitials and split-vacancy complexes as the microscopic origin of the widely observed sub–band-gap emission lines.

\section{Conclusion}

In this work we have carried out a thorough \emph{ab initio} investigation into the possible origins of the below-gap photoluminescence lines of Cu$_2$O. The level of detail required to establish the stability of defect states in a strongly correlated oxide—particularly the analysis of full band structures across multiple supercell sizes and XC-functionals—constitutes, in itself, a new methodology for defect studies within DFT. By applying this approach we find no evidence that the simple copper or oxygen vacancies generate electronic states within the band gap. This implies either that some aspect of the physics of these defects lies beyond what PBE- or HSE‑DFT can capture, or that these vacancies are not responsible for any of the observed PL lines contrary to long-standing assumptions in the literature.

In contrast, the oxygen interstitials consistently produce in-gap states across all functionals and supercell sizes studied, making them strong candidates for at least some of the PL features. The situation for the copper interstitials, the first split copper vacancy and the copper antisite is less clear, since these defects produce mid-gap features under some computational conditions but fail to do so under others. Our method does not allow a quantitative prediction of the PL emission energies so a definitive reassignment of the lines is not possible at this stage. Nevertheless our results support several plausible reassignments involving oxygen interstitials, copper interstitials, and the first split copper vacancy. Future work using excited-state methods such as GW and the Bethe–Salpeter Equation may allow quantitative prediction of defect-related PL energies and thereby test these possibilities directly.

Taken together, these findings provide essential microscopic insight for the growth of high-quality synthetic cuprous oxide and support the development of scalable device technologies based on Rydberg excitons where defect control and accurate interpretation of PL spectra are critically important.

\section{Acknowledgements}
We would like to thank Stefan Scheel and Nikitas Gidopoulos for their generous theoretical discussions. We thank Durham University for HPC resources and also the UK National Facility Archer2 via grant EPSRC EP/X035891/1.


\section{Appendix}
\begin{figure}
    \centering
    \includegraphics[width=0.9\linewidth]{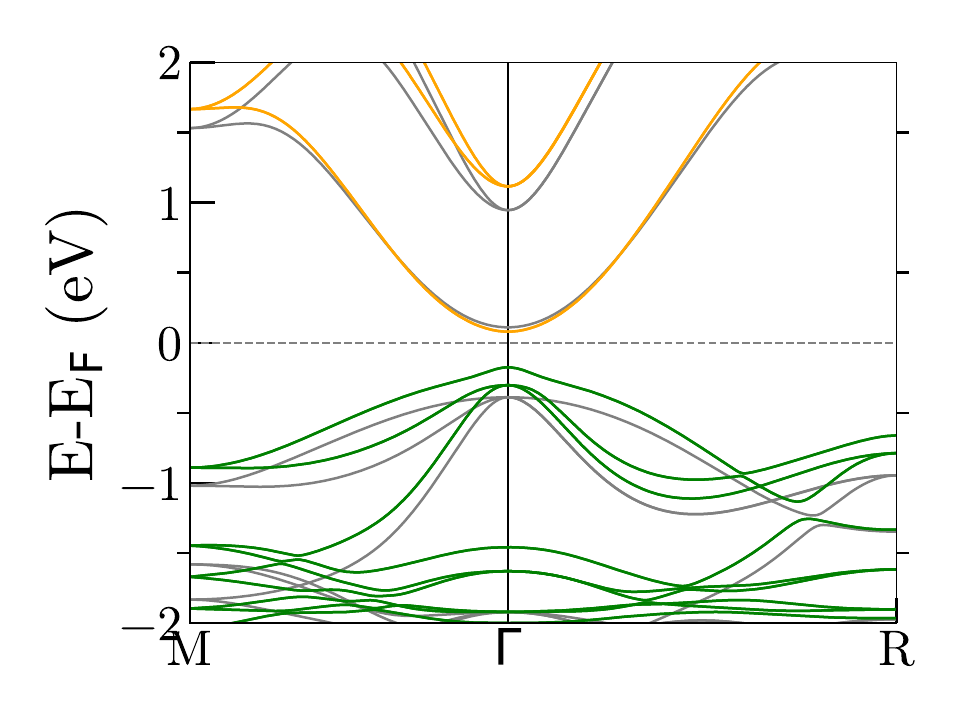}
    \includegraphics[width=0.9\linewidth]{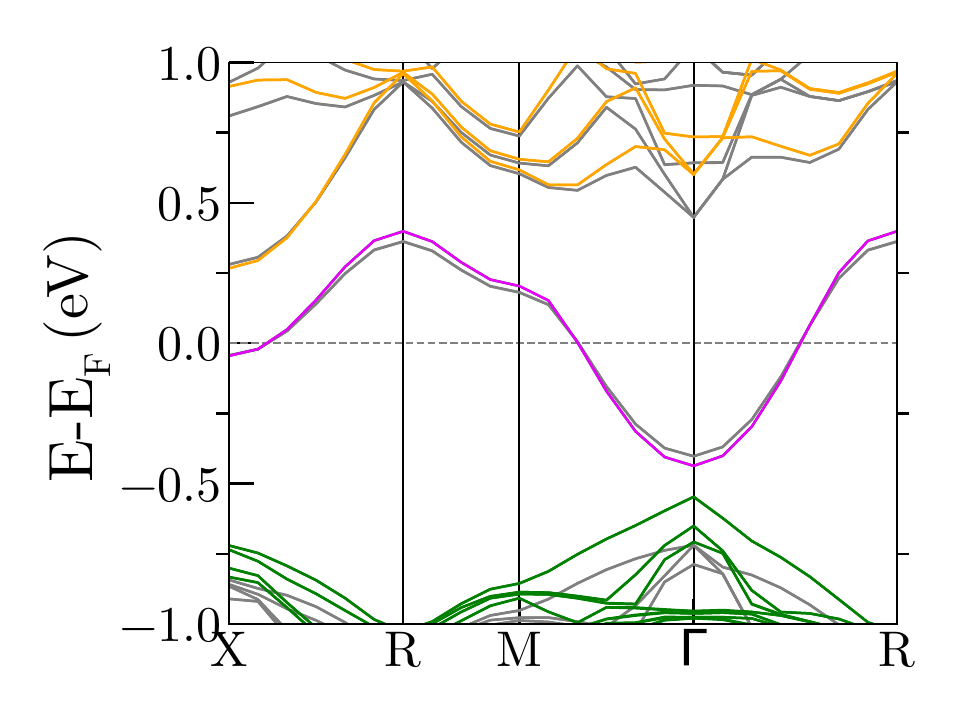}
    \includegraphics[width=0.9\linewidth]{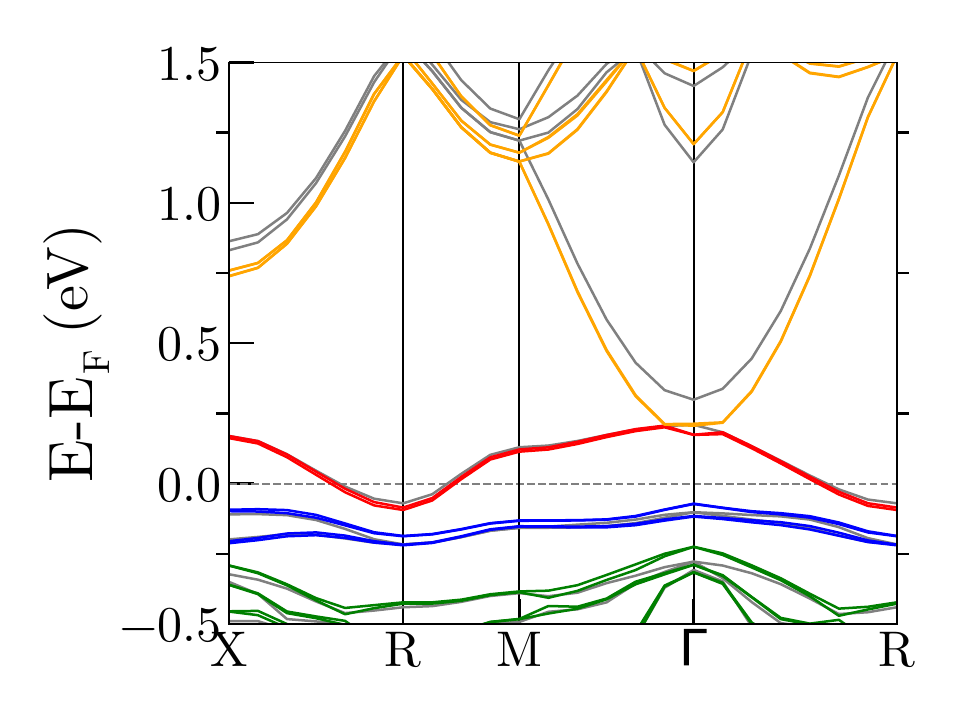}
    \caption{The effect of spin orbit coupling on some example band structures. Top: the Cu$_2$O unit cell. Middle: Cu$_\text{i}^\text{oct}$ under PBE in the $2\times2\times2$ supercell. Bottom: O$_\text{i}^\text{tet}$ under PBE in the $2\times2\times2$ supercell. The band structure without the spin-orbit coupling correction is shown in grey.}
    \label{fig:soc}
\end{figure}

\begin{figure}
    \centering
    \includegraphics[width=0.9\linewidth]{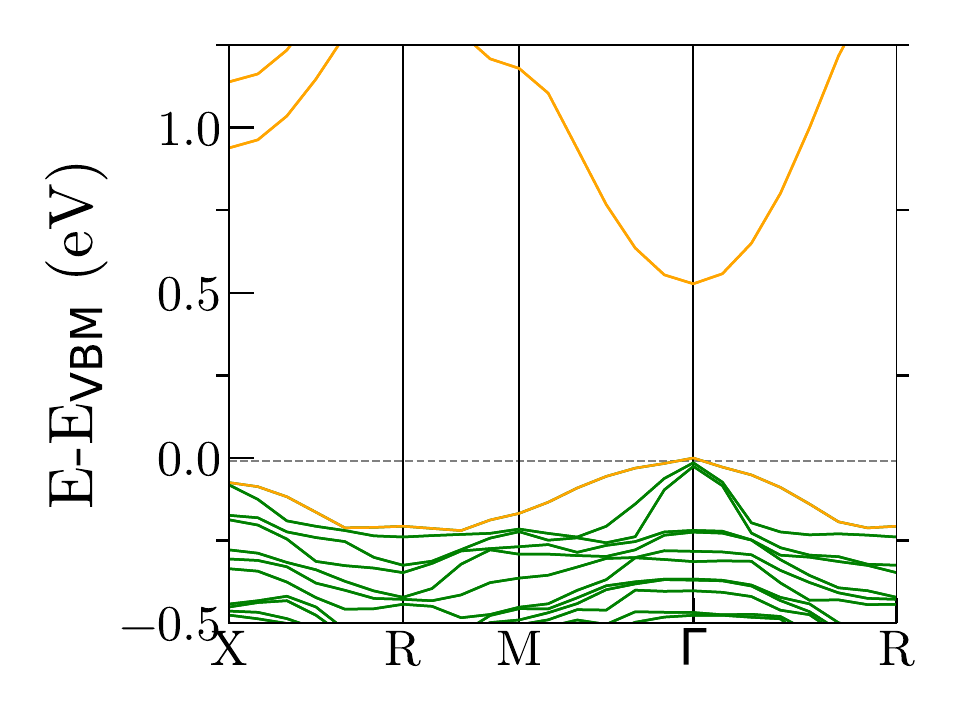}
    \includegraphics[width=0.9\linewidth]{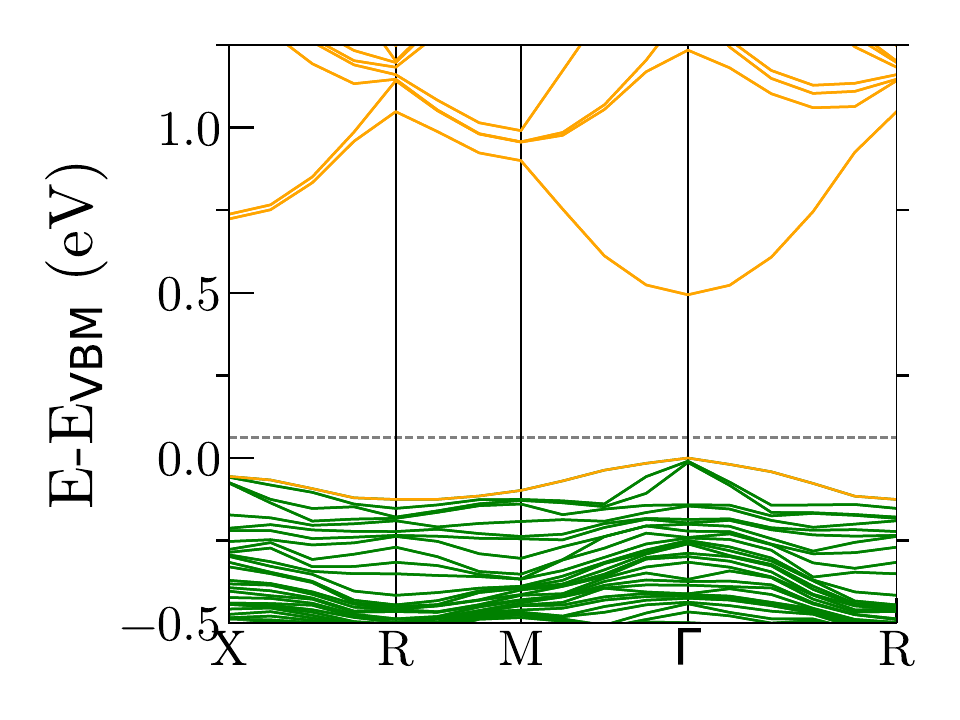}
    \includegraphics[width=0.9\linewidth]{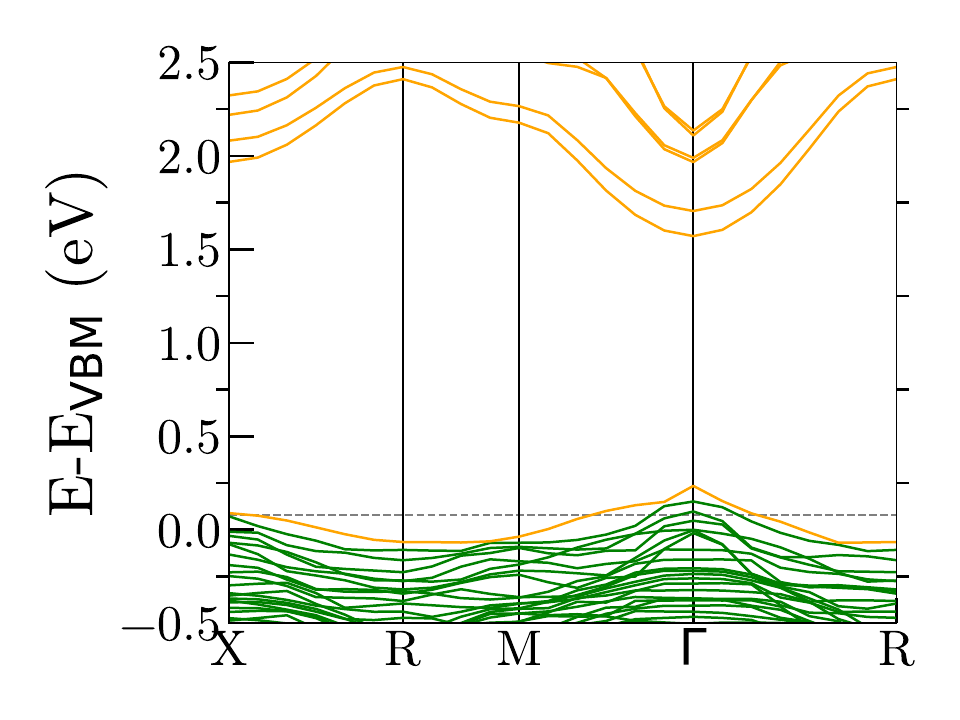}
    \caption{Band structures for the second form of the split copper vacancy, $V_\text{Cu}^\text{s,2}$. From top to bottom: PBE band structure in a $2\times 2\times 2$ supercell; PBE band structure in a $3\times 3\times 3$ supercell; HSE band structure in a $2\times 2\times 2$ supercell. Occupied bulk states are shown in green and unoccupied bulk states are shown in orange. The dotted line denotes the Fermi level.}
    \label{fig:vcus2}
\end{figure}

\begin{figure}
    \centering
    \includegraphics[width=0.9\linewidth]{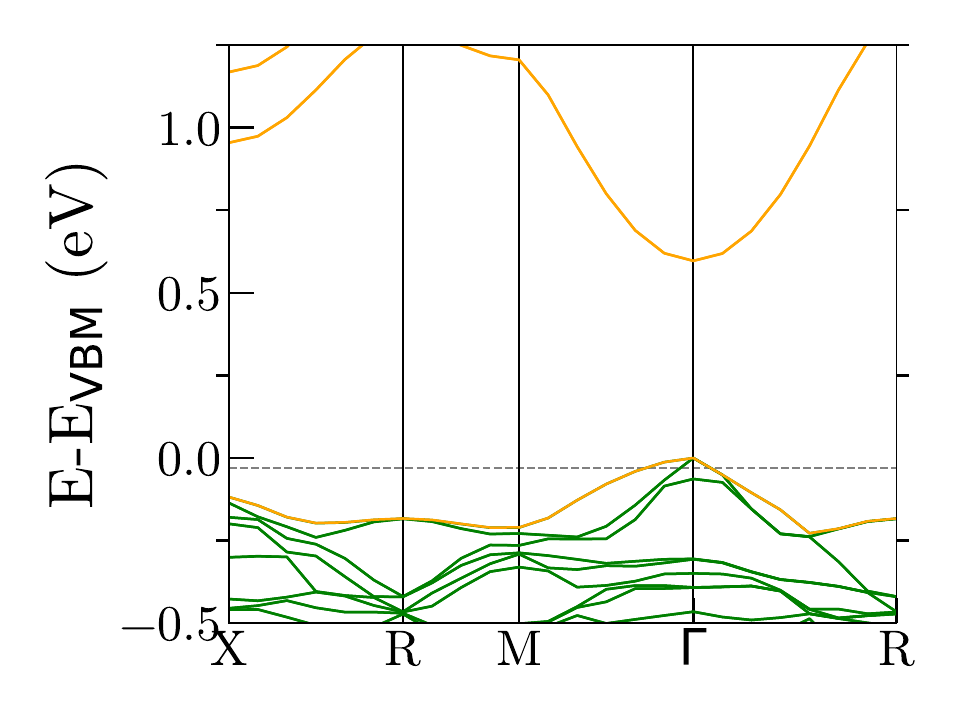}
    \includegraphics[width=0.9\linewidth]{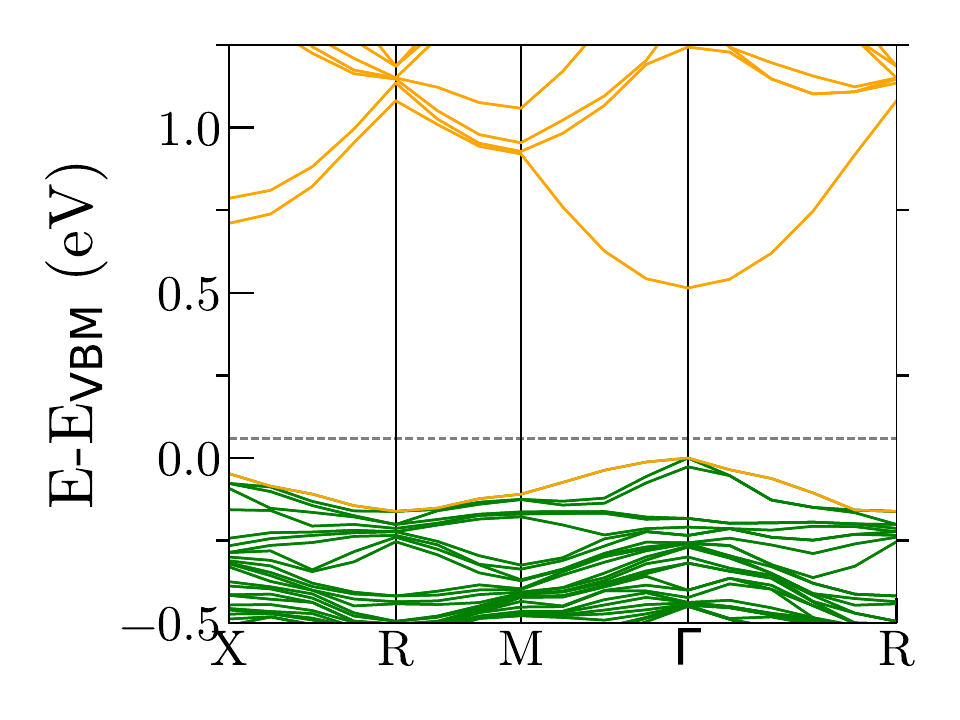}
    \includegraphics[width=0.9\linewidth]{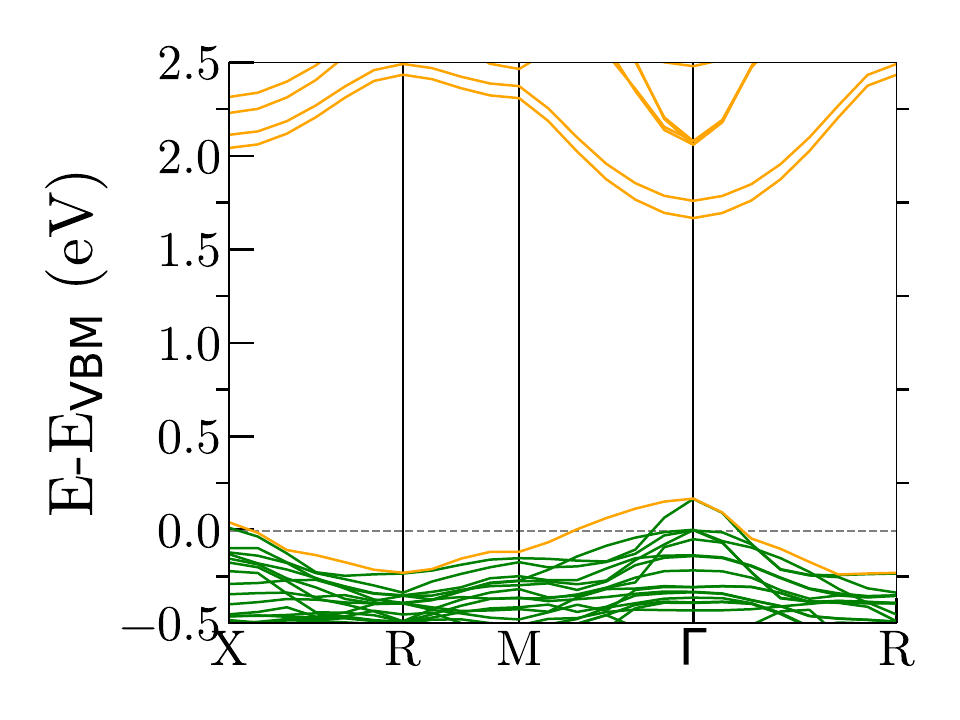}
    \caption{Band structures for the oxygen anti-site, O$_\text{Cu}$. From top to bottom: PBE band structure in a $2\times 2\times 2$ supercell; PBE band structure in a $3\times 3\times 3$ supercell; HSE band structure in a $2\times 2\times 2$ supercell. Occupied bulk states are shown in green, unoccupied bulk states are shown in orange, and unoccupied defect states are shown in red. The dotted line denotes the Fermi level.}
    \label{fig:ocu}
\end{figure}

\begin{appendix}
This appendix contains several band structure plots that were not presented in the main body of the paper to preserve the flow of the text. Namely: examples of the effect of spin-orbit coupling on the unit cell and 2 defect supercells (figure \ref{fig:soc}); the band structures for the second split copper vacancy, V$_\text{Cu}^{s,2}$ (figure \ref{fig:vcus2}); and the band structures for the oxygen anti-site, O$_\text{Cu}$ (figure \ref{fig:ocu}).

\begin{figure}
    \centering
    \includegraphics[width=0.9\linewidth]{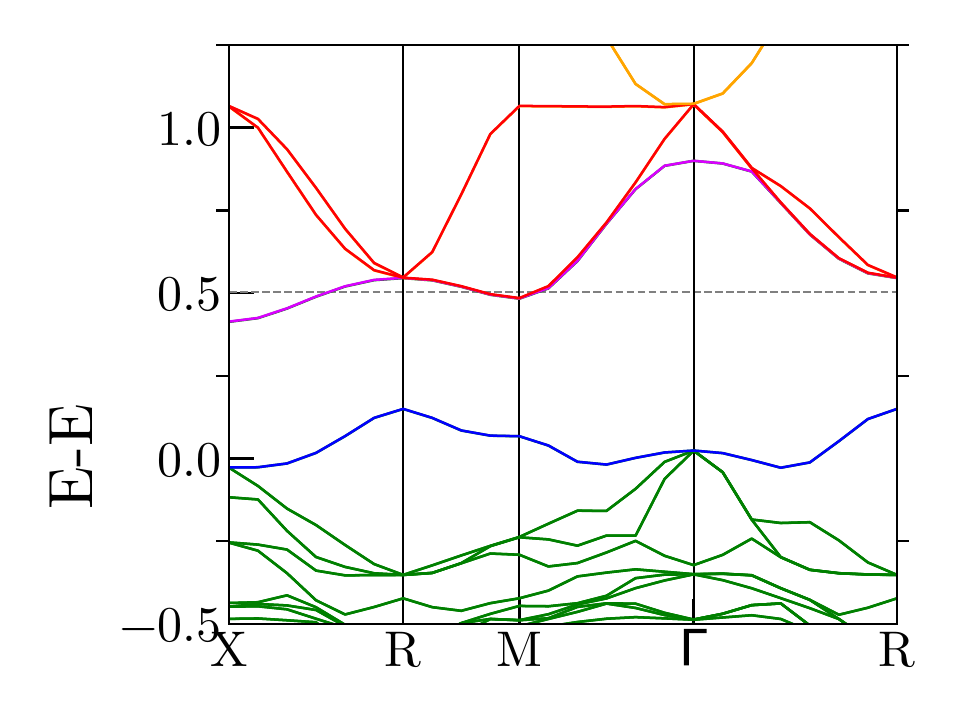}
    \includegraphics[width=0.9\linewidth]{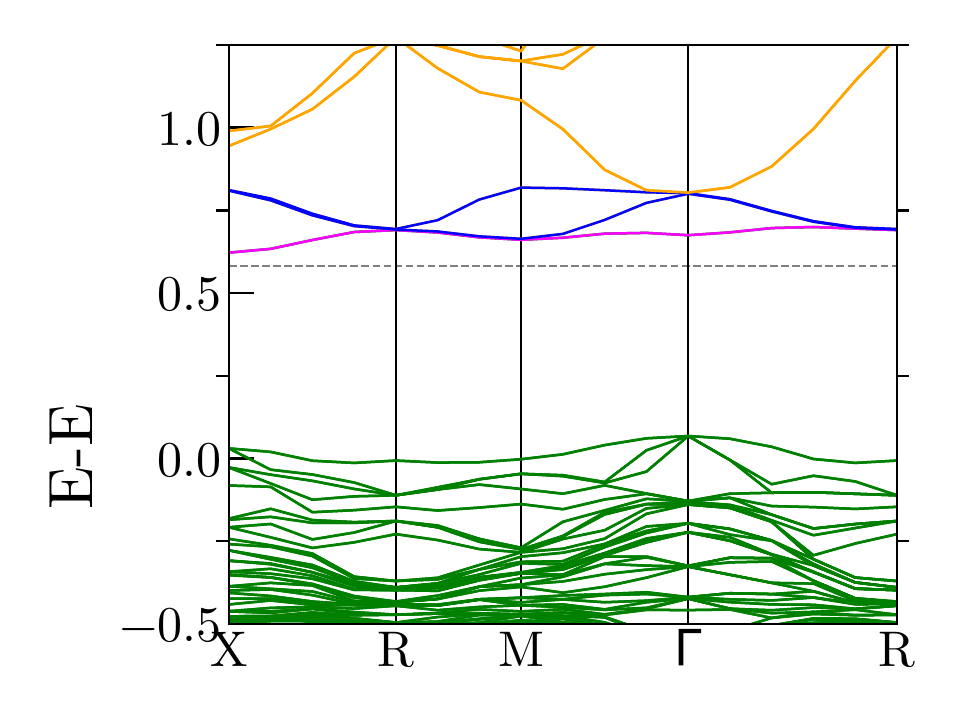}
    \includegraphics[width=0.9\linewidth]{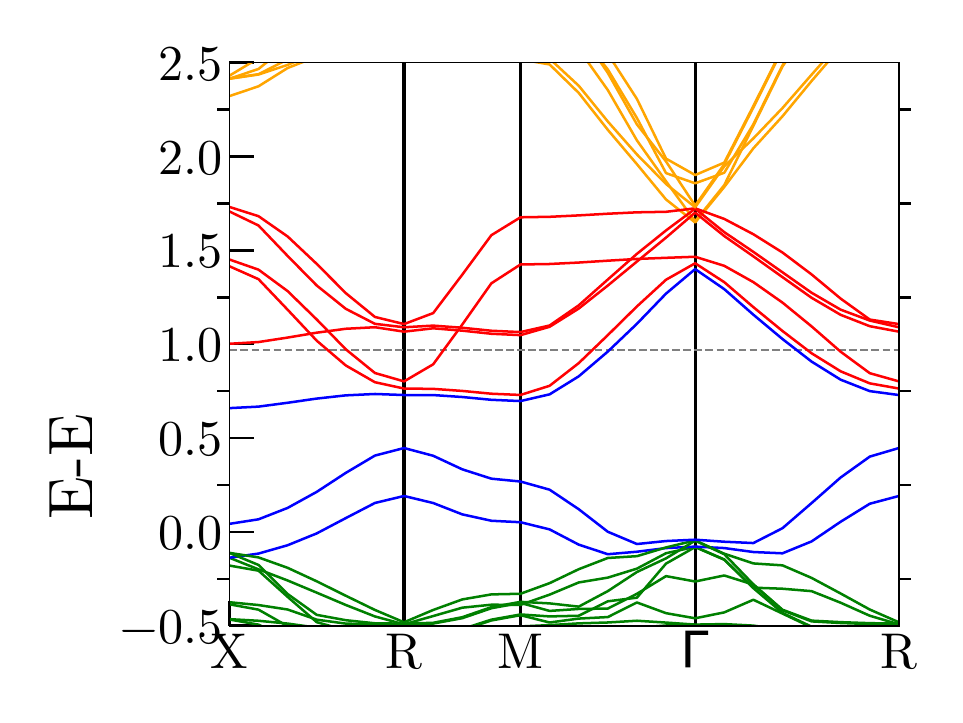}
    \caption{Band structures for the copper anti-site, Cu$_\text{O}$. From top to bottom: PBE band structure in a $2\times 2\times 2$ supercell; PBE band structure in a $3\times 3\times 3$ supercell; HSE band structure in a $2\times 2\times 2$ supercell. Occupied bulk states are shown in green, unoccupied bulk states in orange, occupied defect states in blue, unoccupied defect states in red, and half-filled defect states in purple. The dotted line denotes the Fermi level.}
    \label{fig:cuo}
\end{figure}

\end{appendix}

\bibliographystyle{unsrturl}
\bibliography{main}

\end{document}